\documentclass[twoside]{article}
\pdfoutput=1
\usepackage[a4paper]{geometry}
\usepackage[latin1]{inputenc} 
\usepackage[T1]{fontenc} 
\usepackage{RR}
\usepackage{hyperref}
\usepackage{xspace,xcolor}
\usepackage{amsmath,amsfonts}

\usepackage[lined,ruled,vlined]{algorithm2e}

\usepackage{graphicx}
\DeclareGraphicsExtensions{.pdf}

\if@mathematic
   \def\vec#1{\ensuremath{\mathchoice
                     {\mbox{\boldmath$\displaystyle\mathbf{#1}$}}
                     {\mbox{\boldmath$\textstyle\mathbf{#1}$}}
                     {\mbox{\boldmath$\scriptstyle\mathbf{#1}$}}
                     {\mbox{\boldmath$\scriptscriptstyle\mathbf{#1}$}}}}
\else
   \def\vec#1{\ensuremath{\mathchoice
                     {\mbox{\boldmath$\displaystyle#1$}}
                     {\mbox{\boldmath$\textstyle#1$}}
                     {\mbox{\boldmath$\scriptstyle#1$}}
                     {\mbox{\boldmath$\scriptscriptstyle#1$}}}}
\fi
 
\newcommand{\etal}{{\em et al}.}
\newcommand{\eg}{{\em e.g.}}

\newcommand{\cf}{{\em cf.}}
\newcommand{\ie}{{\em i.e.}}
\newcommand{\siesta}{{\sc siesta}\xspace}
\newcommand{\hypre}{{\sc hypre}\xspace}
\newcommand{\qmmm}{{\sc qm/mm}\xspace}
\newcommand{\tddft}{{\sc tddft}\xspace}
\newcommand{\dft}{{\sc dft}\xspace}
\newcommand{\lda}{{\sc lda}\xspace}
\newcommand{\dzp}{{\sc dzp}\xspace}
\newcommand{\esp}{{\sc esp}}

\newcommand{\fast}{{\sc fast}}
\newcommand{\tsub}[1]{\hbox{\scriptsize #1} }

\newcommand{\dang}{$^{\circ}$}
\newcommand{\wave}{\mbox{$\mathrm{\,cm^{-1}}$}}
\newcommand{\mpicpl}{{\sc mpicpl}}
\newenvironment{acknowledgements}{{\bfseries\noindent Acknowledgements.}\rmfamily\small}{}
\RRNo{8221}
\RRdate{February 2013}
\RRauthor{
Olivier Coulaud\thanks{
Centre de Recherche Inria Bordeaux Sud-Ouest, Olivier.Coulaud@inria.fr}   
\and 
Patrice Bordat\thanks[iprem]{Institut pluridisciplinaire de recherches sur l'environnement et les mat\'eriaux, UMR 5624 du CNRS et de l'Universit\'e de Pau et des pays de l'Adour. 
H\'elioparc, 2, av. du Pt. P. Angot, 64053 PAU Cedex 9,  France.
} 
\and Pierre Fayon \thanksref{iprem} 
\and Vincent LeBris\thanksref{iprem} 
\and Isabelle Baraille\thanksref{iprem} 
\and Ross Brown\thanksref{iprem} 
}
\RRnote{This work was supported by A.N.R. grant "Nossi", CIS-007-005.}

\authorhead{Coulaud }

\RRtitle{Extensions du code \dft\  \siesta\ pour la simulation de mol\'ecules.}
\RRetitle{Extensions of the \siesta\ \dft\ code for simulation of molecules.}
\titlehead{Extensions of the \siesta\ \dft\ code}
\RRresume{Nous d\'ecrivons les extensions au code siesta~\cite{Soler2002} de la th\'eorie de la fonctionnelle de densit\'e (\dft),
pour la simulation des mol\'ecules isol\'ees et leurs spectres d'absorption. Ces extensions permettent  : 
\begin{itemize}
\item{l'utilisation d'un solveur multigrille pour l'\'equation de Poisson sur le maillage \dft\.  Les conditions aux limites de Dirichlet sont calcul\'ees par un d\'eveloppement en harmoniques sph\'eriques du potentiel \'electrique ;}
\item{la coupure du syst\`eme mol\'eculaire \`a l'aide du pseudo-potentiels de l'atome sur mesure de Xiao Zhang\cite{Xiao2007} ;}
\item{le calcul des charges effectives atomiques par la m\'ethode de l'ajustement du potentiel \'electrostatique ; }

\item{Calcul des \'energies de transition d'absorption \'electroniques et des forces d'oscillateur \`a partir des spectres bruts obtenus par un code \dft\ d\'ependant du temps\cite{Koval2010}.
Le code est en outre int\'egr\'e dans \siesta\ comme une option de post-traitement. code\cite{Koval2010}.}
\end{itemize}
}
\RRabstract{
We describe extensions to the \siesta\ density functional theory (\dft)
code~\cite{Soler2002}, for the 
simulation of $isolated$ molecules and their absorption spectra. The extensions allow for:
\begin{itemize}
\item{Use of a multigrid solver for the Poisson equation on a finite \dft\ mesh. Non-periodic, Dirichlet boundary conditions are computed by expansion of the electric multipoles over spherical harmonics.}
\item{Truncation of a molecular system by the method of design atom
pseudo-potentials of Xiao and Zhang\cite{Xiao2007}.}
\item{Electrostatic potential fitting to determine effective atomic charges.}
\item{Derivation of electronic absorption
transition energies and oscillator strengths from the raw spectra produced by a
recently described, order $O(N^3)$, time-dependent \dft\ code\cite{Koval2010}. 
The code is furthermore integrated within \siesta\ as a post-processing option.}
\end{itemize}
}
\RRmotcle{solveur multigrille, calculs \dft\  /\tddft\, syst\`eme mol\'eculaire, \siesta\ }
\RRkeyword{multigrid solver, \dft\ /\tddft\ computation, molecular systems, \siesta\ }
\RRprojet{HiePACS}  

\RCBordeaux 
\begin{document}
\makeRR   

\section{Introduction}
\label{intro}
The \siesta\ program is well established in the field of simulation of solids
with density functional theory (\dft)\cite{Soler2002}. It is used in an ever
widening range of applications, benefitting from regular maintenance and
extensions of the code's capabilities\cite{Artacho2008,Sanz-Navarro2011}.
\siesta\ employs numerical atomic orbitals (AO's) with strictly finite range,
leading to order-$N$ scaling of computations with respect to the number of
atoms. \siesta\ was thus an attractive \dft\ engine for a new, fast
time-dependent \dft\ (\tddft) algorithm for molecular systems, based on use of
dominant products of finite orbitals, and scaling as $O(N^3)$ with the
number of atoms $N$\cite{Koval2010}. Previously, this \tddft\ code used orbital
and density matrix data from files produced by \siesta. Moreover, it provided
only a raw spectrum. We have therefore undertaken to couple the program
directly to \siesta\, and to extract transition
energies and oscillator strengths from the raw spectrum. Furthermore, \siesta\
remains a periodic \dft\ code, meaning that to simulate a molecular system,
very large, essentially empty \dft\ meshes may be necessary to effectively
isolate the system from its periodic images.  

The present contribution therefore describes extensions of \siesta\ useful in the field of molecular systems:
\begin{enumerate}
\item{Adaption to molecular computations by  {\em (i)} Computation of Di\-ri\-chlet
bound\-ary con\-di\-tions in the Poisson equation, by development of electric
multipolar moments, up to order 4 in spherical harmonics; and  {\em (ii)} Introduction of a multi-grid solver for the
Poisson equation;}
\item{Introduction of an electrostatic potential fit algorithm for the
assignment of atomic partial charges;}
\item{Implementation of  'design atom' pseudo-potentials\cite{Xiao2007},
allowing truncation of a molecular system by replacing a bond by a tailor-made
lone pair;}
\item{Direct coupling to \siesta\ of the order  $O(N^3)$ \tddft\ code \fast, with
extraction of transition energies and oscillator strengths.}
\end{enumerate}

Extensions 1--3,discussed in the correspondingly numbered subsections of
part \ref{modifs}, are available as a set of patches of \siesta, development version 431,
downloadable at 

 \href{https://gforge.inria.fr/frs/?group\_id=1179}{https://gforge.inria.fr/frs/?group\_id=1179}\footnote{Access to be opened on acceptance of the present paper}. The 
\fast\ \dft\ code is available at the same {\sc url}.

\section{Extensions of \siesta}
\label{modifs}
\subsection{Solution of the Poisson equation for non-periodic systems \label{poisson}\label{Non-periodic}}

During self-consistency cycles in a \dft\ computation, recourse is made at each
cycle to the Hartree energy,
\begin{equation}
E_H=\int \rho(\vec{r})V(\vec{r}) d\vec{r}   , \label{Hartree}
\end{equation}

\noindent approximated as a discrete sum over a mesh of points encompassing the
system. The electrostatic potential, $V(\vec{r})$, itself is formally a space
integral of contributions of the electronic density, $\rho(\vec{r})$,
\begin{equation}
V(\vec{r})=\int \rho(\vec{s})\frac{1}{\|\vec{r}-\vec{s}\|} d\vec{s}   . \label{IntegralV}
\end{equation}
\noindent Rather than directly integrating the contributions of infinitesimal
volume elements of the density, it is much more efficient to solve the Poisson
equation for $V(\vec{r})$:

\begin{equation}
\Delta V(\vec{r}) = -4\pi\rho(\vec{r})/\epsilon_{0} , \label{Poisson}
\end{equation}

\noindent with suitable boundary conditions, where $\Delta$ is the Laplacian.  We refer the reader to \cite{Soler2002} for details of implementation in \siesta, in particular
partition of the problem into neutral atom and bond contributions to $\rho$.

\begin{figure}
\begin{minipage}[t][1\totalheight]{1\textwidth}
\begin{center}
\includegraphics[scale=.4, angle=0]{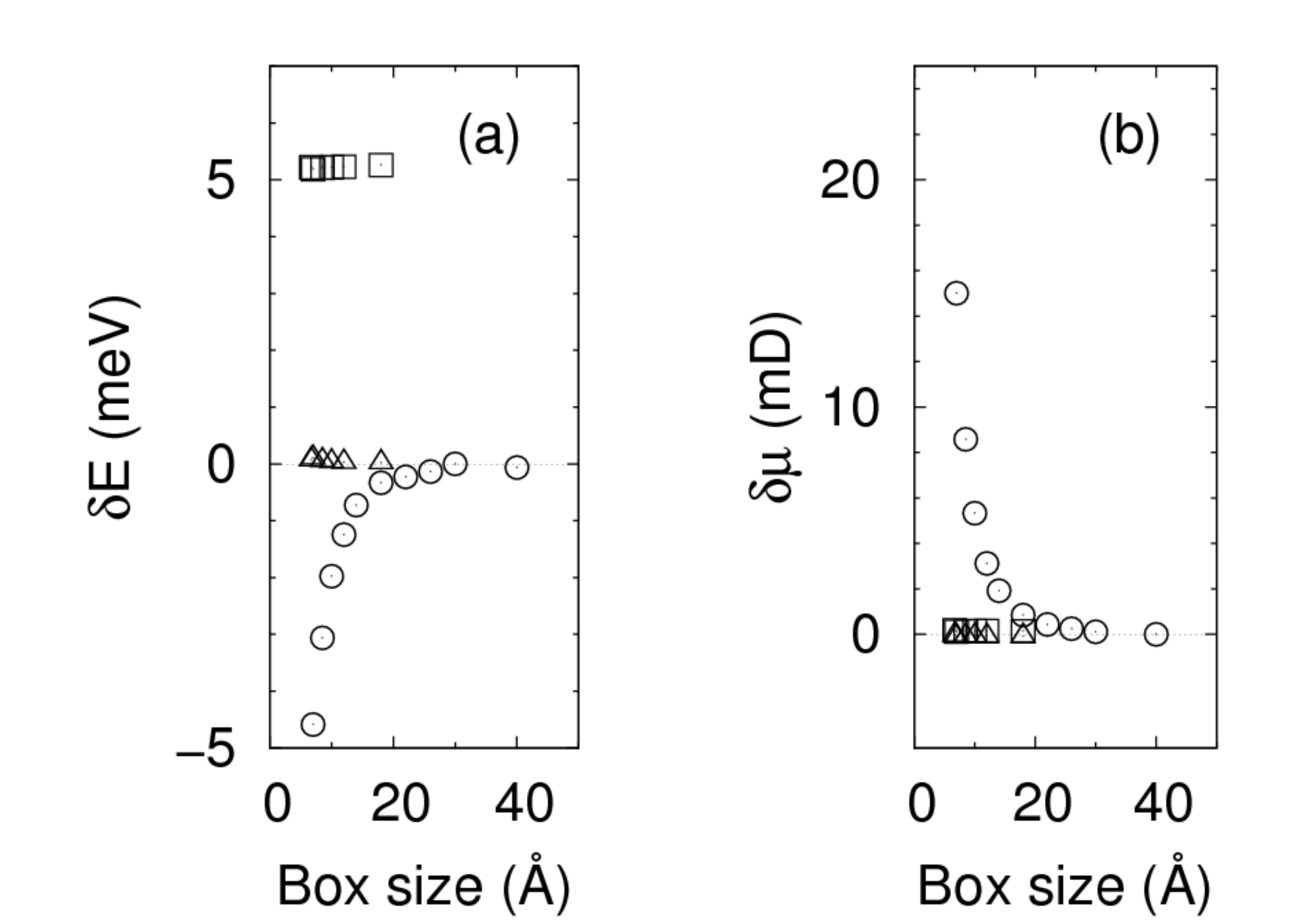} 
\end{center}
\end{minipage}
\caption{Convergence of the physical properties of the water molecule with
respect to the side of the simulation box: {\em (a)} Energy; {\em (b)} Electric dipole moment. Symbols : $\circ$, periodic simulation, using the
standard FFT solver;  $\Box$ and $\triangle$, non-periodic simulation, using
a second order, 7-point  or a fourth order, compact 16-point stencil respectively 
for the Laplacian in the multi-grid solver for the Poisson
equation.  Data are plotted as differences with respect to the
converged values (FFT, large box). See section \ref{Non-periodic} for simulation
conditions. 
\label{H2Oprops}} 
\end{figure}

Because \siesta\ is a periodic code, in which the system and the \dft mesh
(typically one crystal unit cell) are replicated indefinitely in all
directions, a particularly effective solution of eqn.  (\ref{Poisson}) is
obtained by transforming to $\vec{k}$-space using fast Fourier transforms
(FFT). This method is ideal for crystals, but for finite molecular systems the
\dft mesh (periodic cell) must be made large enough to damp out interactions of
the system with its periodic images. Physical properties may converge only
slowly with the mesh size. Figure (\ref{H2Oprops}a,b) illustrates this for a toy
model of an isolated water molecule. Although the molecule and its orbitals
hold within a cube of side $5.7$\,\AA\,, a \dft mesh of side 20\,\AA\ is
required to achieve good convergence of the energy ($\approx-465.8$\,eV) and
dipole moment ($\approx 1.389$\,D) in the periodic FFT computation.  

We refer repeatedly to this model, using the standard double-zeta with
polarisation basis set of \siesta\ (\dzp, energy shift parameter 0.02\,Ry), in
the local density approximation (\lda, Ceperley-Alder
exchange-correlation\cite{Perdew1981}). The mesh cutoff is 400\,Ry. Heuristics
in \siesta\ cause the corresponding mesh step to vary slightly with the box
size, in the range $0.08 \pm 0.002$\,\AA\ in all calculations. The tolerances
for convergence of the density matrix and energy are $10^{-6}$ and
$10^{-6}$\,eV respectively.  Standard Troullier-Martins pseudo-potentials from
the \siesta\ library are used throughout. The model is used at the optimised
geometry. 

In fact, the solution of eqn. (\ref{Poisson}) for finite systems is determined uniquely
 by the values on a closed surface enclosing the charge
density, \ie\ Dirichlet boundary conditions.  The \dft mesh for a finite system can
thus be made quite small, provided there is a recipe for accurately specifying
the solution on the surface of the \dft mesh.

A molecular computation thus needs; {\em (i)} an independent means of
specifying the Dirichlet boundary conditions at the surface points of the
finite \dft mesh;  {\em (ii)} an efficient solver for the boundary value problem
of $V(\vec{r})$ in the interior of the domain.

\paragraph {Dirichlet boundary conditions by expansion of multipoles over
spherical harmonics.}
Dirichlet boundary conditions may be determined by either of two alternative
multipole expansions of $\frac{1}{\|\vec{r}-\vec{s}\|}$ in
eqn~(\ref{IntegralV}): Cartesian multipoles or spherical harmonics. Cartesian
expansion in the inverse distance\cite{Jackson1998} is easy and cheap to
compute up to second order, $p=2$, but tedious to pursue to higher orders.
Furthermore, the number of terms grows as $p^3$. Spherical harmonics cost more
to evaluate, but are readily extended to any order. The number of terms grows
only  as $(p+1)^2$ and their complexity is lower than in the Cartesian
approach.  We have implemented expansion over spherical harmonics in \siesta.
The electrostatic potential on the surface of the simulation box is
approximated by an expansion up to order $p$ as follows\cite{Greengard1987}:
\begin{eqnarray*}
V(\vec{r}) \approx V_p(\vec{r}) & = & \sum_{l=0}^p \sum_{m=-l}^l \frac{(-1)^m}{r^{l+1}}M^m_l Y^{-m}_l(\theta,\phi) , \\
M^m_l & = & \int_{\mathbb{R}^3} d^3\vec{s} \rho(\vec{s})s^l Y^{-m}_l(\vec{s}) ,
\end{eqnarray*}
\noindent where $(r,\theta,\phi)$ are the spherical coordinates of point $\vec{r}$. In practice, the electronic density $\rho$ is known on the \dft mesh
points, and non-vanishing only within the supports of the finite numerical
orbitals employed in \siesta. We allow only orthorhombic meshes for the
multi-grid solver, since an arbitrary mesh is pointless for computation of an
isolated molecule.  Therefore, although we develop the potential over spherical
harmonics on the boundaries we express them in Cartesian form. 
%
Setting
$\phi(\vec{s})=s^l Y^{-m}_l(\vec{s})$ at mesh point $\vec{s}$, we
approximate $M^m_l$ by
\begin{equation*}
M^m_l \approx \sum_{\vec{s}|\rho(\vec{s})\ne 0} \rho(\vec{s}) \phi(\vec{s}) dV ,
\end{equation*}

\noindent  where $\phi(\vec{s})$ is computed only at mesh-points with
non-vanishing density, \ie\  above a user-defined threshold. The volume
element of the mesh is  $dV$.

The multipole expansion converges formally for all $\vec{r}$ outside the
support of $\rho(\vec{s})$.  Accurate Dirichlet conditions may be obtained
either by: increasing the size of the \dft mesh box, improving in eqn.
(\ref{IntegralV}) the separation between surface points $\vec{r}$  and charges
at interior points $\vec{s}$; or by increasing the order of the expansion. In
\dft computations, intensive use of the mesh points drives the balance towards
higher order expansion on a smaller mesh.

If the supports of the numerical orbitals impinge on the surface of
the \dft mesh, accuracy is certainly compromised. 
In our implementation, we use the atomic coordinates, and the radii of the
atomic orbitals to check for this problem. However, it is always advisable
to perform a series of exploratory calculations with different box sizes and
multipole orders, to prove convergence of computed properties with respect to
errors in the expansion.

As can be seen from figure \ref{H2Oprops}, both the total energy and the
dipole moment of water are highly converged when boundary conditions are
computed with fourth order spherical harmonics in  a box of only $7$\,\AA\,,
or a clearance of barely 0.65\,\AA\ around the atomic orbitals used to develop
$\rho(\vec{r})$.

\paragraph {Multi-grid solver for the Poisson equation.}
Solution of the Poisson equation on a regular finite mesh proceeds by
discretizing the Laplacian operator by finite differences.  One obtains a system of linear equations,
$\vec{A}_h \vec{V} = \vec{f}$, for $V$ at points in the interior of the domain. Vector $\vec{f}$
contains the Dirichlet conditions. The sparse, symmetric, positive definite
matrix $\vec{A}_h$  has the classical stripe pattern described by a 'stencil'.
A 7-point second order stencil and a compact, 16-point fourth order solver have been included 
in the present implementation. The matrix size is $N=N_1
\times N_2\times N_3$ where $N_i=L_i/h_i$ is the number of mesh points in the
$i$-direction and $h_i$ is the mesh step in $i$-direction.   

Iterative methods are attractive for solving the Poisson equation inside the
SCF loop, rather than direct methods like Cholesky factorization.
It is well known that multi-grid solvers are the fastest iterative methods for
solution of the Poisson equation in a rectangular box. The complexity is linear
in the system size, even better than for a periodic solver based on FFT.
Reference\cite{Briggs2000} discusses multigrid methods (MG) in detail.  One
efficient parallel multigrid software package is the {\sc hypre} software library
\cite{hypreDesign,hypreManual,hypre}. Here, we wrap this general library to make it available within
\siesta.  We use the structured-grid interface and 
either PFMG, a semicoarsening multigrid solver that uses pointwise relaxation, or preconditioned conjugate gradients (PCG)\cite{hypreSolver}. The PFMG solver allows
different parallel smoothers, including the red-black Gauss-Seidel method.  The PGC method uses PFMG as a preconditioner. Our wrapper is very flexible and can be extended easily to more complex operators
like higher order discretizations of the Laplacian.  Different kinds of mesh
distribution are also available. We use classical, uniform 2-D real-space domain
decomposition and the new \siesta\ parallelization scheme based on balanced
real-space domain decomposition, achieved with a recursive bisection
algorithm\cite{Sanz-Navarro2011}. Ideally, an irregular or even  unstructured mesh would adjust to the
gradient of the electronic density, leading to a finite element formulation of
the problem, a future refinement possible within the \hypre\ library.

\paragraph{Performance.}

\begin{figure}
\begin{center}
\includegraphics[scale=.6]{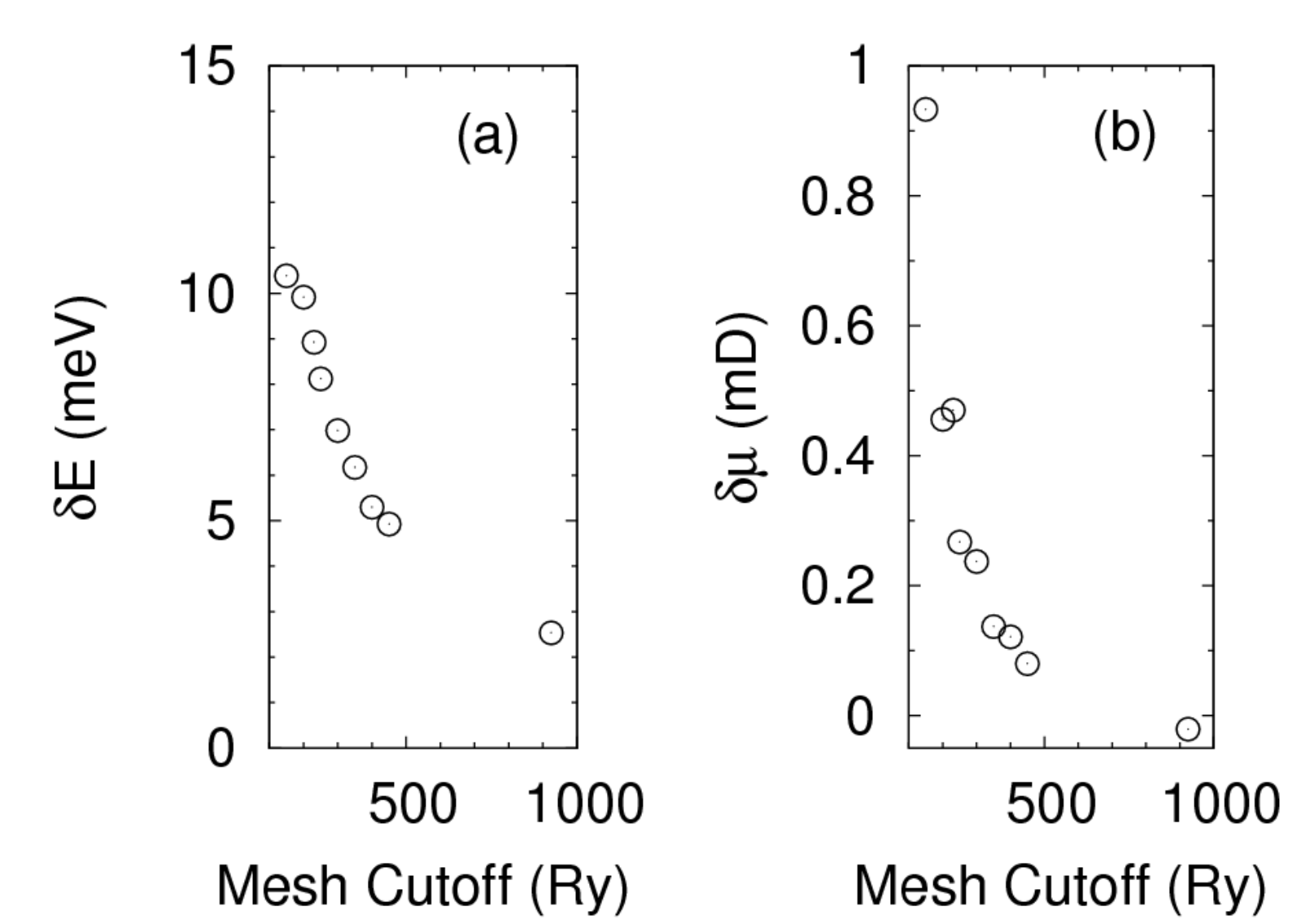} 
\end{center}
\caption{Dependence of the properties of the water molecule on the mesh step in
the finite model with the second order MG solver, box side  7\,\AA\,. Data are shown as
differences with respect to the well converged FFT solution of a large periodic
model (box side 40\,\AA). (a) Energy(ref. -465.800453\,eV); (b) Dipole
moment (ref. 1.3887\,D). \label{MGmeshdependence}} \end{figure}

Implementation of the {\sc hypre} MG solver was checked by generating the
Dirichlet conditions for distributions of point charges, numerically solving
the Poisson equation in empty space and checking the solution against the known
analytical result. The embedding of the MG solver in \siesta\ was checked by
comparing to results with the standard FFT in \siesta, on the toy model of
water mentioned above. In what follows we found it
sufficient to expand the Dirichlet boundary conditions over spherical
harmonics up to order four.

Figure (\ref{H2Oprops}a,b) shows, for the same mesh step 
in the FFT and MG calculations (0.08\,\AA\ or plane wave cutoff of 400\,Ry),  the superior
convergence of the physical properties with the finite box size, achievable
with the MG solver and Dirichlet conditions, compared to that of the solution
under periodic boundary conditions, with the FFT solver. For the given mesh
step, the energy obtained with the second order MG solver is $5$\,meV higher than the FFT
solution, and the dipole moment 0.1\,mD larger. Indeed, the final accuracy of the
\dft\ computation depends on the step of the DFT mesh and on the order of
discretization of the operator.  The energy obtained with the second order
discrete Laplacian operator is converged with respect to
box size, but to a value significantly different from the FFT result. The fourth order approximation, using a
16-point stencil, on the other
hand, gives the same results as the FFT calculation. Figure (\ref{MGmeshdependence}a,b) confirms convergence of the MG
solutions to the FFT values as the step size is reduced, for the second order solver, where the effect is most significant.
Figure 1 of the supplementary information(SI) shows the linear scaling of the algorithm with the system size.

\subsection{Electrostatic potential fit of partial atomic charges}
\label{ESP}

Often it is desirable to assign charges to atomic centres, to be used as
proxies for the full electronic density $\rho(\vec r)$. Such schemes, \eg\
Mulliken charges and 'atoms in molecules' analysis\cite{Terrabuio2012}, each
have their own advantages and disadvantages. In classical molecular dynamics,
partial atomic charges 'best fit' to represent the Coulomb forces would be
desirable. In practice, 'best fit' representation of the Coulomb potential itself is
much more tractable analytically and more common than fitting the gradient of
the potential.

We therefore have added such an electrostatic potential (\esp) fit routine to
\siesta. This is a classic
problem\cite{Bayly1993,MillerFrancl1996,Sigfridsson1998}. Here we give the
minimum details to describe our implementation.  The problem is to determine a
set of partial atomic charges $q_i, i=1,\ldots,N$ on $N$ atomic centres
$\vec{R}_i$, such that their Coulomb potential is a good approximation to the
full molecular potential at a set of test points $\vec{S}_j,
j=1,\ldots,M$, commonly chosen near the molecular van der Waals surface.  Let
$U_j$ be the full molecular electrostatic potential at $\vec{S}_j$:
\begin{equation*}
U_j = -\int_{\mathbb{R}^3} d^3\vec{r} \frac{\rho(\vec{r})}{\vert\vec{r}-\vec{S}_j\vert} + \sum_{i=1}^N \frac{ Z_i}{\vert\vec{R}_i-\vec{S}_j\vert} ,
\end{equation*}
\noindent where $Z_i$ is the nuclear charge on atom $i$ (in the case of pseudo-potentials, the
valence charge). These values are to be approximated by
\begin{equation*}
V_j=\sum_{i=1}^N \frac{ q_i}{\vert\vec{R}_i-\vec{S}_j\vert} , 
\end{equation*}
\noindent where the partial charges $q_i$ are determined by least squares
minimisation of the error 
\begin{equation*}
\chi^2\left(\{q_i\}\right)=\sum_{j=1}^M \left(U_j -V_j\right)^2 .
\end{equation*}
\noindent As usual, we add {\em via}\ a Lagrangian multiplier, $\lambda$, the
constraint that the $q_i$ should sum to the total charge of the molecule, $Q$.
The problem is then to find stationary points (minima) of the Lagrangian
\begin{equation*}
L\left(\left\{q_i\right\},\lambda\right)= \chi^2\left(\{q_i\}\right) + \lambda \left(\sum_{i=1}^N q_i -Q\right).
\end{equation*}

\def\matrixAp
{\begin{array}{cccc}
    &         &      & 1            \\
    & \vec{\Sigma\Sigma}^{\tsub{T}} &      & \vdots       \\
    &         &      & 1            \\
1   & \ldots  &   1  & 0
\end{array}}
\def\qpvec
{\begin{array}{c}
 \vdots            \\
 q_i       \\
 \vdots     \\
 \lambda
\end{array}}
\def\bpvec
{\begin{array}{c}
 \vdots            \\
 \sum_{j=1}^M U_j/s_{ij}     \\
 \vdots     \\
 Q
\end{array}}
\noindent After a little algebra one finds the $q_i$ (and multiplier $\lambda$)
as solutions of  an $(N+1) \times (N+1)$ linear system, the matrix equation
\begin{equation}
\vec{A}'\vec{q}'=\vec{b}' ,
\label{Apqpbp}
\end{equation}
\noindent with
\begin{equation*}
\vec{A}' =  \left( \matrixAp \right),\qquad \vec{q}' =  \left( \qpvec
\right),\qquad \vec{b}' =  \left( \bpvec \right) . 
\end{equation*}
\noindent  $\vec{\Sigma}$ is the NxM matrix with elements $\Sigma_{ij}=1/s_{ij}$, with $s_{ij}=\vert\vec{R}_i-\vec{S}_j\vert$.

It long has been known\cite{MillerFrancl1996} that matrix $\vec{A}'$ may be
ill-conditioned or singular, depending on the choice of the test points
$\vec{S}_{j}$, because the far field of a set of charges is determined
principally by their lowest order multipoles. In our implementation we use as
$\vec{S}_{j}$'s a subset of the \dft\ mesh points, since the full electrostatic
potential is known already on the \dft\ mesh at the end of the SCF cycles.
Similar to a Connolly surface\cite{Connolly1983}, the mesh points chosen are those
at least at a distance $R_{\tsub{min}}$ from all nuclei, and at most $R_{\tsub{min}}+\delta R$
from some nucleus (the 'skin' thickness $\delta R$ being in practice
$\approx 0.1$\,\AA). Our experience  with the finite support AO's in \siesta\ is that
\esp\ charges are physically sensible and relatively insensitive to $R_{\tsub{min}}$
when it is chosen from just greater than the largest AO radius, up to around
$5$\,\AA\ beyond. For larger values, matrix $\vec{A}'$ may indeed 
become singular. Typical mesh sizes generally lead to several thousand points
within the skin, so that the main cause of any indetermination of the charges
is the singularity of $\vec{A}'$ when all the points are chosen too far away
from the molecule.

\begin{figure}
\begin{minipage}[t][1\totalheight]{1\textwidth}
\begin{center}
\includegraphics[scale=.4]{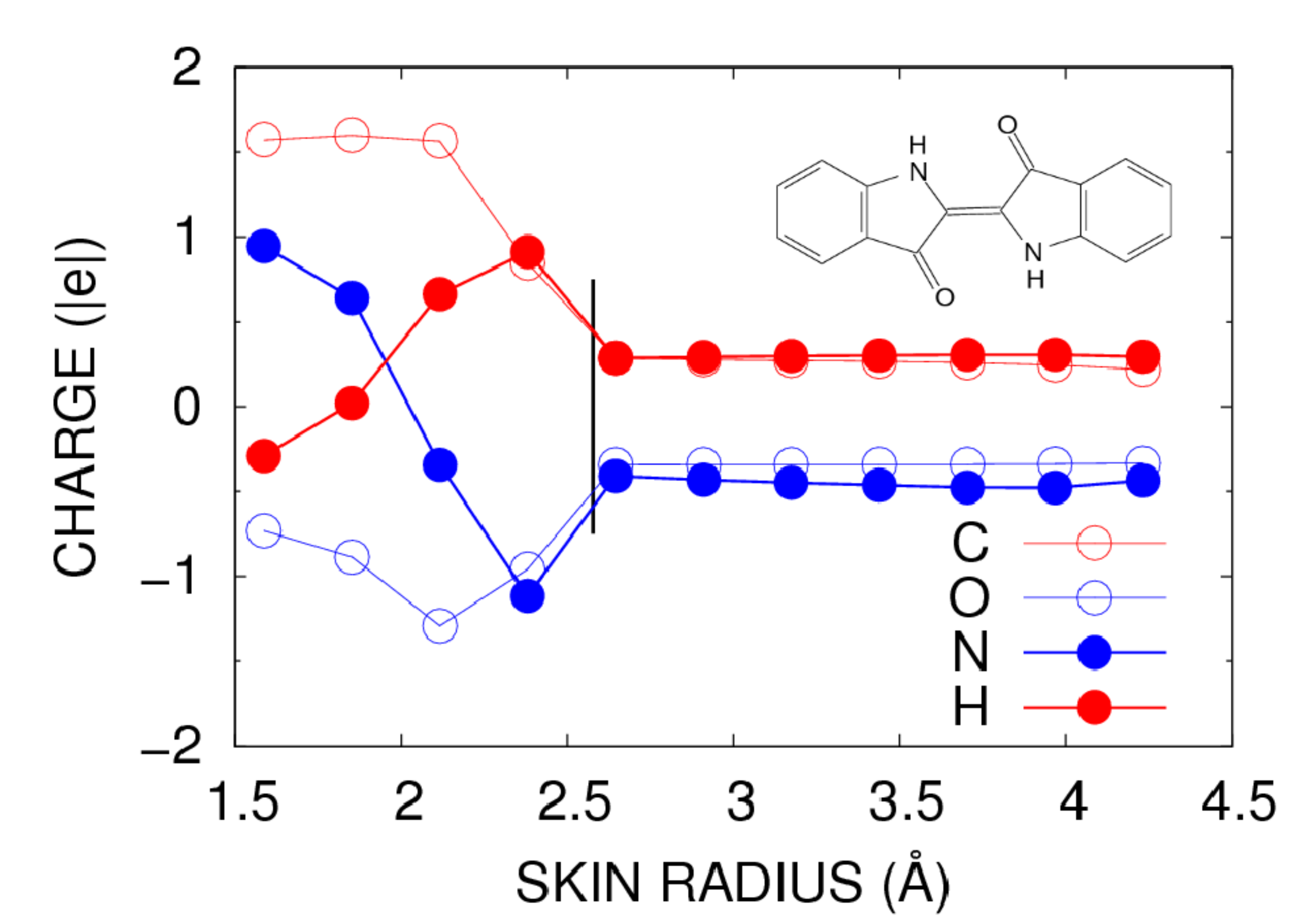} 
\end{center}
\end{minipage}
\caption{ \esp\ charges on the  polar groups C=O and NH in indigo (inset), as a
function of the radius of the 'skin' of \dft\ mesh points used in the \esp\
fit. The vertical line shows the largest atomic orbital radius.   \label{charges}}
\end{figure}

We solve equation (\ref{Apqpbp}) using a truncated SVD algorithm, starting with $QR$ factorization of
$\vec{A}'$ as $\vec{A}'=\vec{Q}\vec{R}$, where $\vec{Q}$ is an orthogonal matrix and
$\vec{R}$ is upper triangular. Singular value decomposition of $\vec{R}$ yields
$\vec{R}=\vec{U}\vec{W}\vec{V}^T$, where $\vec{U}$ and $\vec{V}$ are orthogonal
and $\vec{W}$ is diagonal (diagonal elements $w_i$). Small singular eigenvalues,
signifying singularity of the matrix and indetermination of the $q_i$, are
removed by setting $w_i=0$ if $\left| w_i/w_{\tsub{max}} \right| < \epsilon$.
We standardly choose $\epsilon=10^{-9}$ . 
The solution, $\vec{q}'$, of eqn. (\ref{Apqpbp})  is then
\begin{equation*}
\vec{q}' = \vec{V} \vec{W}^{-1} \vec{U}^T \vec{Q}^T \vec{b}'.
\end{equation*}

\noindent Routines from the {\sc blas}\cite{blas} and {\sc lapack}\cite{lapack} libraries are used here to perform these operations. The calculation is parallel.

Figure (\ref{charges}) shows the \esp\ charges found for polar groups in indigo,
as a function of the \esp\ shell radius $R_{\tsub{min}}$ ($\delta R = 0.1 $\,\AA\,).
Conditions are the same as for the toy water model in section
\ref{Non-periodic}.  The Poisson equation was solved with the multi-grid method with
Dirichlet boundary conditions from spherical harmonics up to order 4. It will
be observed that the \esp\ charges are insensitive to $R_{min}$ once the sample
points are all outside the largest atomic orbital (radius $\sim 2.6$\,\AA). The charges on C,O,N and H
are 0.28, -0.34, -0.42 and 0.29e\,.  

\subsection{Truncation of molecules with design atom pseudo-potentials\label{design}
}
\begin{figure}
\begin{minipage}[t][1\totalheight]{1\textwidth}
\begin{center}
\includegraphics[scale=.6]{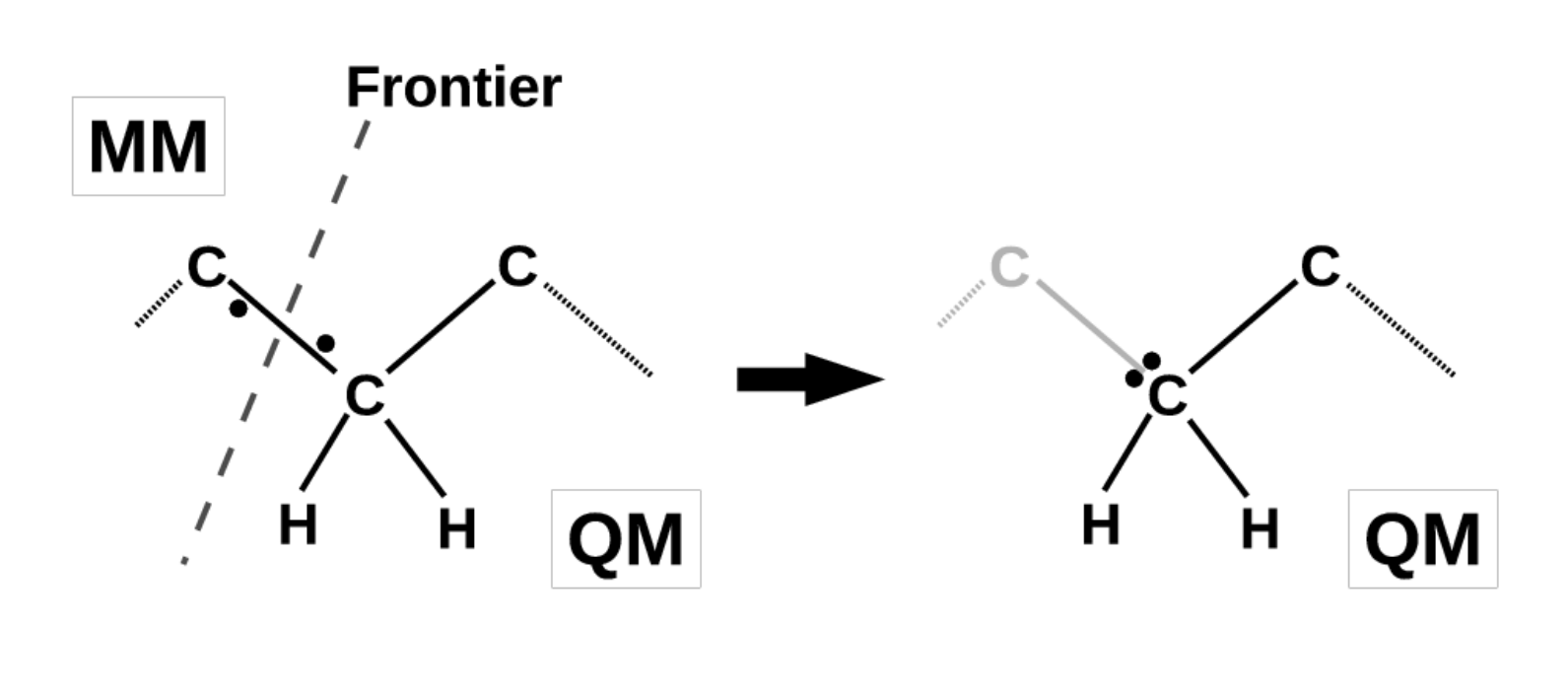} 
\end{center}
\end{minipage}\caption{Truncation of a large molecule in a QM/MM model by
cutting a bond and completing the dangling valence with an electron to form
a lone pair mimicking the missing bond.\label{lonepair}}
\end{figure}

It often is desirable to restrict the size of a quantum mechanical calculation
to just the essential atoms for the problem at hand. Quantum
mechanics/molecular mechanics (\qmmm) calculations are a prime example, as are
cluster calculations in which a cluster of atoms of tractable size is cut out
of a periodic material. In both cases it is necessary to correct dangling
valencies left by the truncation, to reduce perturbation of the core region of
the cluster. One way is to add capping atoms (usually hydrogens) to complete
peripheral valencies.

Another is to turn dangling bonds into lone pairs, an approach developed by
Xiao and Zhang\cite{Xiao2007} with a view to \qmmm\ calculations, \cf\ figure
(\ref{lonepair}).  Perturbation of the core of the cluster is then minimised by
transforming the peripheral atom so capped into a 'design' atom, with specific
pseudo-potentials to mimic the electronic structure of the atom it replaces.
We have extended this method to truncation on oxygen and applied it to larger systems, where
we find the perturbation decays rapidly with distance from the cut bond. 
\begin{figure}
\begin{minipage}[t][1\totalheight]{1\textwidth}
\begin{center}
\includegraphics[scale=.4]{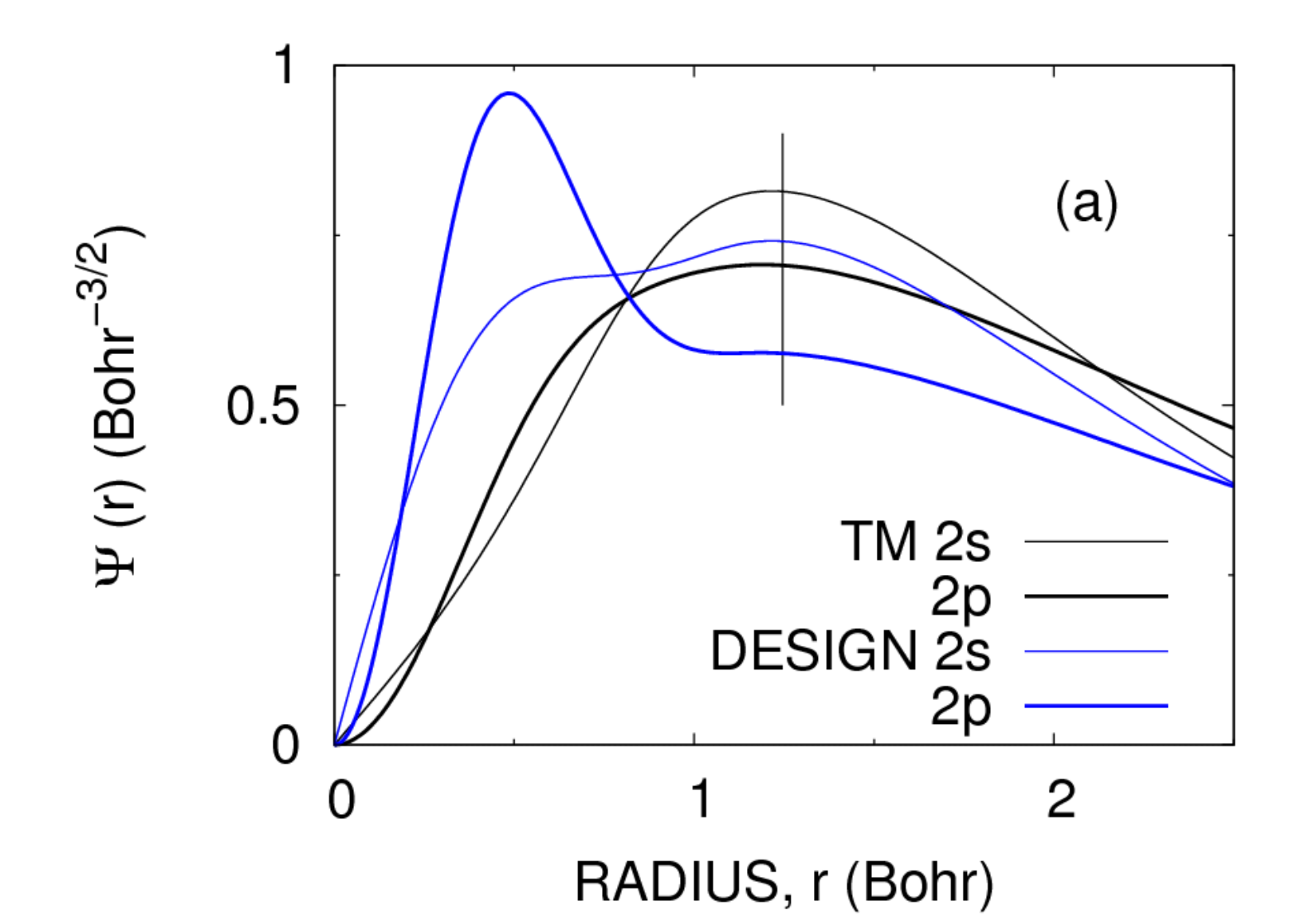} 
\end{center}
\end{minipage}
\begin{minipage}[t][1\totalheight]{1\textwidth}
\begin{center}
\includegraphics[scale=.4]{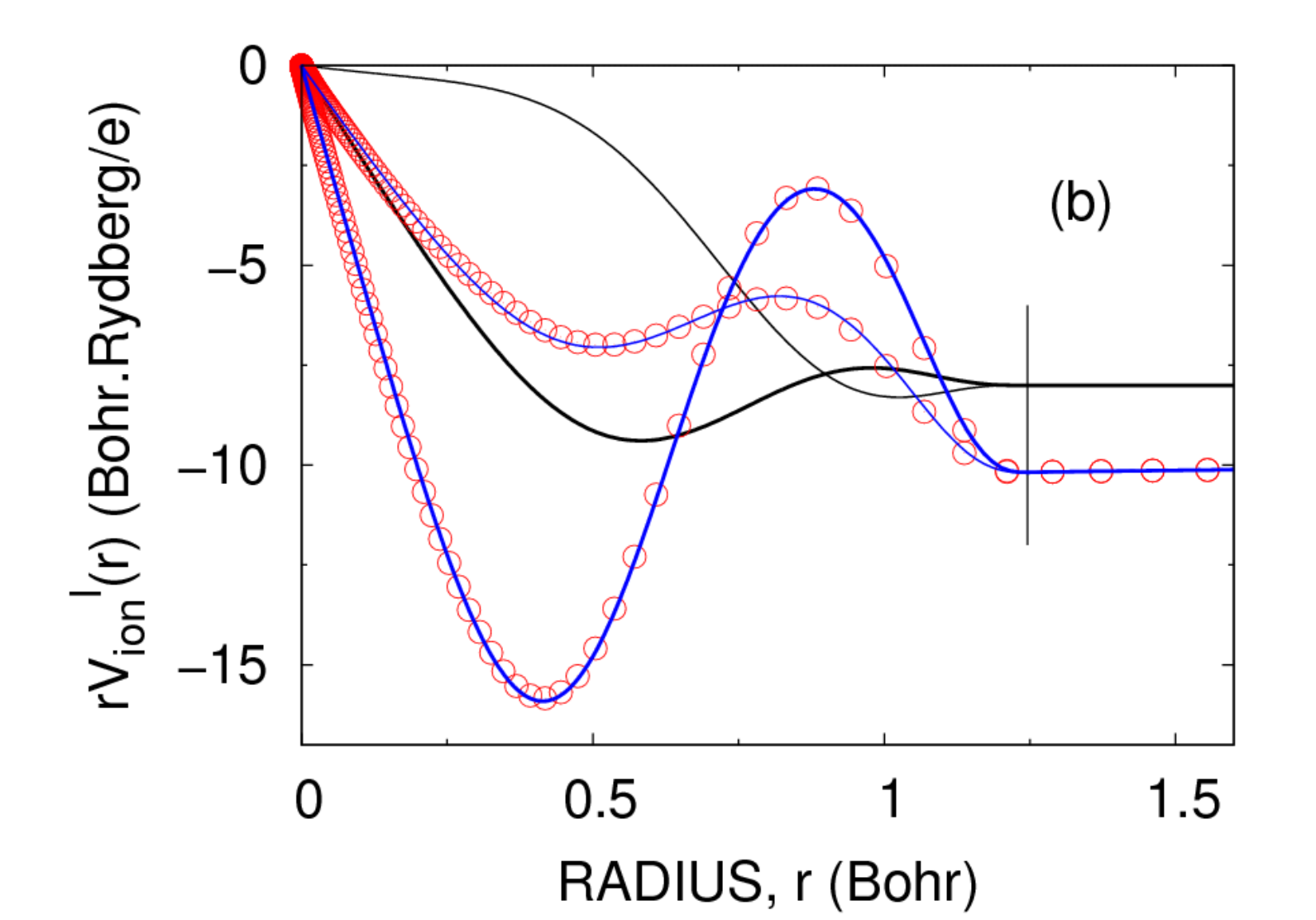} 
\end{center}
\end{minipage}
\caption{  Comparison of design atom and standard Troullier-Martins
pseudo-potentials for the 2s (thin) and 2p (thick lines) shells of carbon.
Black: standard Troullier-Martins method, blue: design atom method, this work.
Vertical lines: core radius. {\em (a)} Wave functions, showing squeezing from
outside to inside the core to maintain the valence density of standard carbon;
{\em (b)}\  Ionic pseudo-potentials, showing deepening in the core to
accommodate the extra electronic density. Points: reference
data kindly provided by Y. Zhang. \label{designpseudos}}
\end{figure}

Pseudo-potentials may be produced and tested with the help of the {\sc atom}
program distributed with  \siesta. We coded 'design' Troullier-Martins, norm
conserving pseudo-potentials in {\sc atom}. In the design atom approach for
carbon, one adds an electron and increases the nuclear charge by one unit. But
the design atom is not just nitrogen, since to minimise discrepancies between the full and the truncated molecule, it is further required that the electronic density of the
design atom outside the core matches that of carbon.  This constraint is
achieved by rescaling the wave functions outside the core radius, per angular
momentum channel, $l$, by $\eta_l = \left(N_l/N^D_l\right)^{1/2}$, where $N_l$
is the number of valence electrons of the original atom in shell $l$ and
$N^D_l$ that in the same shell of the design atom.
\begin{figure}
\begin{minipage}[t][1\totalheight]{1\textwidth}
\begin{center}
\includegraphics[scale=.25]{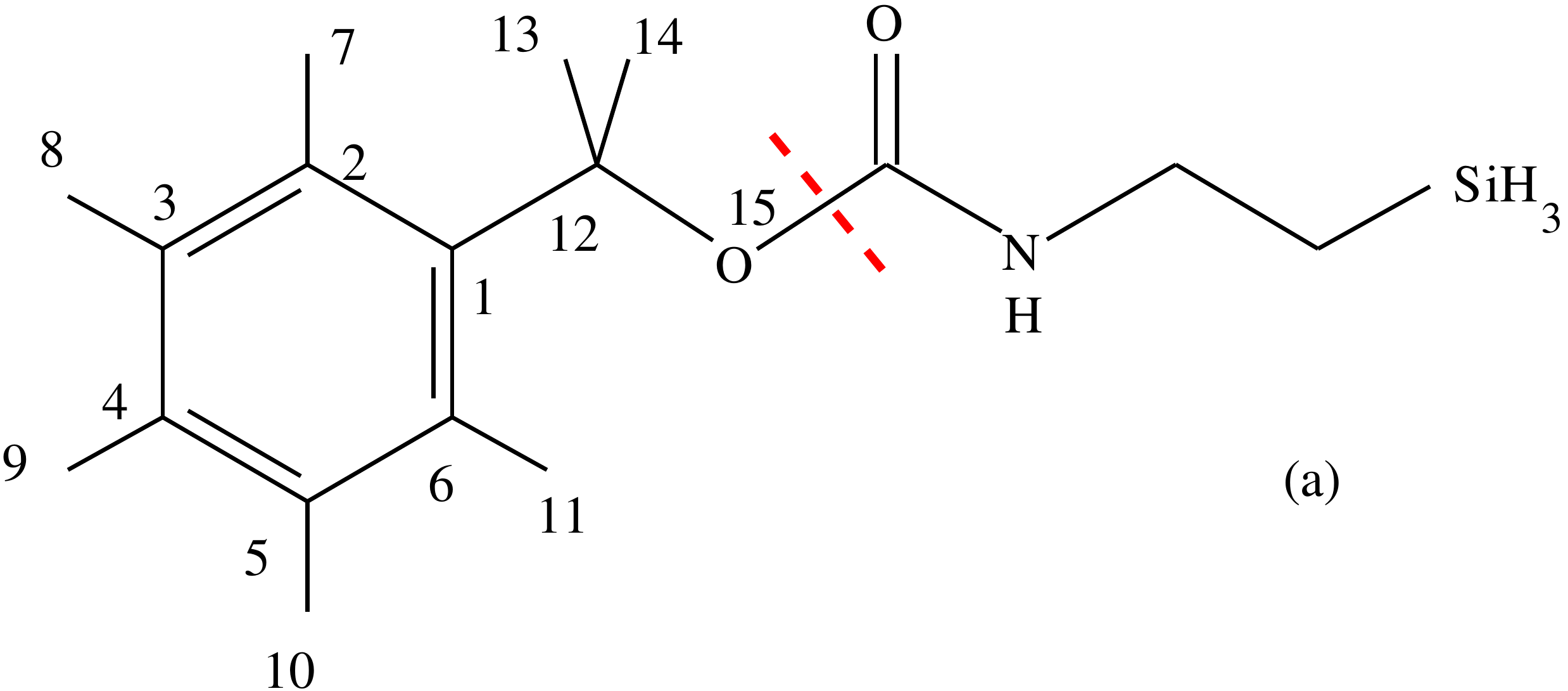} 
\end{center}
\end{minipage}
\begin{minipage}[t][1\totalheight]{1\textwidth}
\begin{center}
\includegraphics[scale=.3]{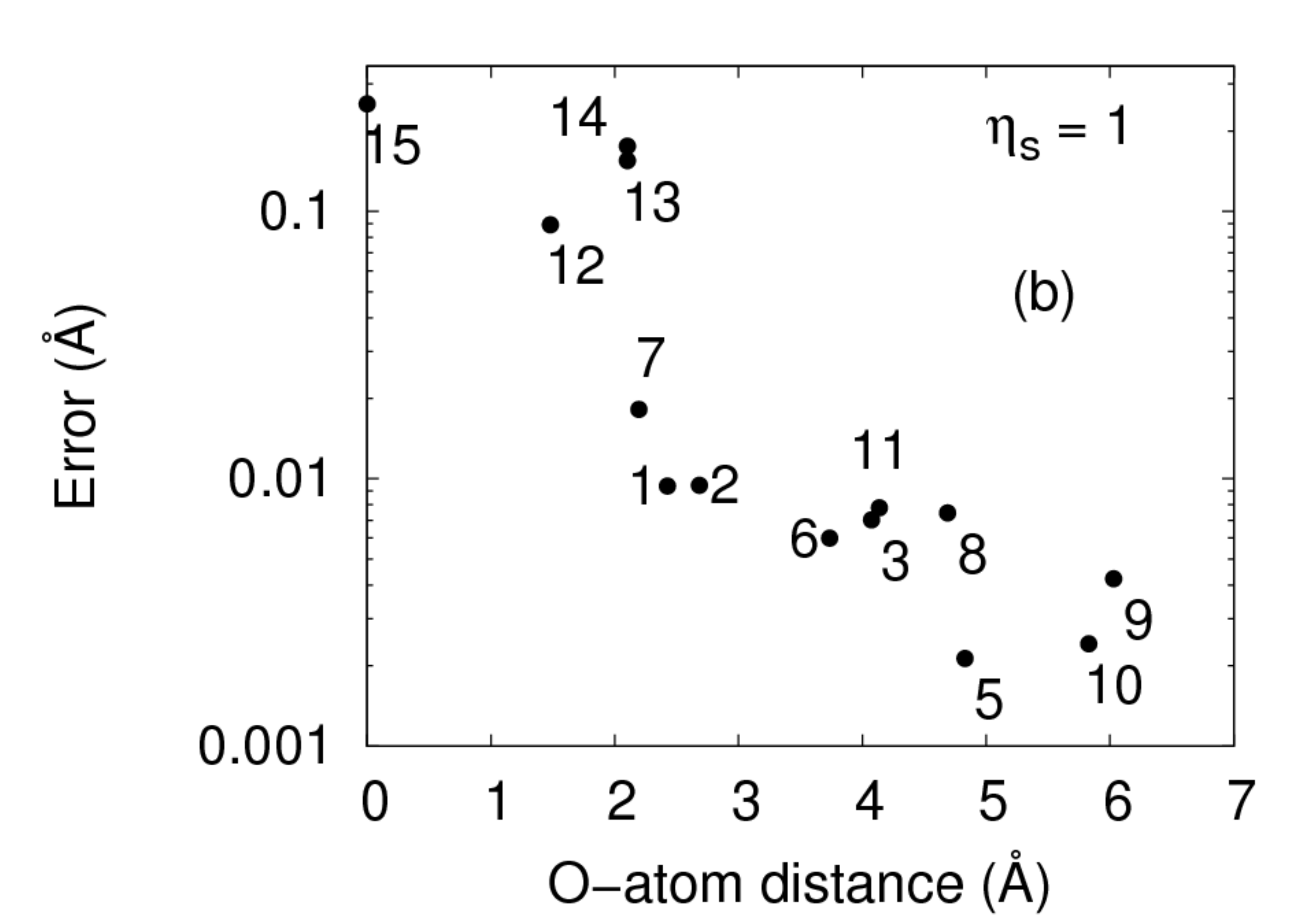} 
\end{center}
\end{minipage}
\caption{ Use of a design atom oxygen to truncate a benzyl carbamate model of a
graftable dye: {\em (a)} Structure, indicating the bond cut and replaced by a lone pair;
 {\em (b)}\  Errors in the positions of atoms in the
truncated molecule with respect to the full molecule (numbering as in {\em (a)}).  \label{benzylmachin}}
\end{figure}

Thus, for carbon, $\eta_{\hbox{\scriptsize 2p} }=\sqrt{2/3}$, and one expects
$\eta_{\hbox{\scriptsize 2s} }=1$. Compared to  standard carbon, the radial  2p
valence wavefunction of design carbon is therefore  depressed outside the core
radius, and exalted within, to accommodate the extra electron, figure
(\ref{designpseudos}a). In fact Xiao and Zhang adjusted
$\eta_{\hbox{\scriptsize 2s}}$, optimising it with respect to the geometrical
parameters of their target molecules, yielding in their case
$\eta_{\hbox{\scriptsize 2s}} =0.91$.  Our implementation closely follows ref.
\cite{Xiao2007}, and indeed our design carbon pseudo-potentials agree very
closely with those of Xiao and Zhang, see figure (\ref{designpseudos}b).

We extended the design atom approach to truncation on oxygen. In this case,
$\eta_{\hbox{\scriptsize 2p} }=\sqrt{4/5}\sim 0.89$. We  varied
$\eta_{\hbox{\scriptsize 2s} }$ in a series of truncations on the oxygen atom
of the toy benzyl carbamate in figure (\ref{benzylmachin}a), similar in
structure to graftable photosensitizers of reactive oxygen species of interest
to us\cite{Lacombe2009}. Here, we show results for $\eta_{\hbox{\scriptsize 2s}} =1$;
reducing it deteriorated the results. Figure (\ref{benzylmachin}b) compares the geometries of
the full and the truncated forms, both fully optimised (\lda, standard \dzp\
basis). The important part of the molecule is the phenyl ring, standing in for
the chromophore. Our measure of quality is therefore to bring three phenyl
carbons (atoms 4,1 and 3 in fig. \ref{benzylmachin}a), of both the full and the
truncated molecules, respectively to the origin, on to the $Ox$ axis and into
the $Oxy$ plane, and to compute the distances between corresponding atoms,
shown in fig.  (\ref{benzylmachin}b) as a scatter plot of error {\em vs}.
distance to the oxygen atom before truncation. It will be observed that the
error of placement of atoms in the fragment relative to the full molecule, drops
off fast as a function of the distance from the design oxygen, being under
10$^{-2}$\,\AA\ for those in the ring.  Complete $Z$-matrices are provided in the supplementary information.

\subsection{Linear response \tddft}
\label{tddft}

\subsubsection*{Context}
A new, order  $O(N^3)$ \tddft\ code was described
recently~\cite{Foerster2008,Foerster2009,Koval2010}, exploiting the strictly
finite range of the numerical atomic orbitals in \siesta. The optical
absorption of a molecule, for a light electric field at angular frequency $\omega$, polarised in
the $\alpha$ direction ($\alpha=x,y,z$), is proportional to the complex part of
the polarisability
\begin{equation}
P_\alpha(\omega)= \int d^3\vec{r} d^3\vec{r}' \vec{r}_{\alpha}
 \vec{\chi}\left( \vec{r}, \vec{r}',\omega \right)\vec{r}_{\alpha}' ,
\end{equation}
\noindent where $\vec{\chi}\left( \vec{r}, \vec{r}',\omega \right)$ is the
susceptibility, and we use the fact that molecules are much smaller than the
wavelength. The susceptibility  $\vec{\chi}$ is expressed in terms of that of
the non-interacting electrons in the Kohn-Sham approach, $\vec{\chi}_0$, and
the Hartree and exchange-correlation kernel $\vec{\Sigma}$:
\begin{equation}
\vec{\chi}\left( \omega \right) = \left(  1- \vec{\chi}_0\left( \omega
\right) \vec{\Sigma} \right)^{-1}\vec{\chi}_0\left( \omega
\right) .    \label{chitot}
\end{equation}
\noindent The non-interacting electron response function $\vec{\chi}_0$, reads
in terms of Kohn-Sham orbitals:
\begin{eqnarray}
\vec{\chi}_0\left( \vec{r}, \vec{r}',\omega \right) & =&
\sum_{E<0,F>0} \psi_E(\vec{r})\psi_F(\vec{r}) \psi_F(\vec{r}')\psi_E(\vec{r}')\nonumber \\
& & \times \left( \frac{1}{\omega -(E-F) + j\epsilon} -
  \frac{1}{\omega + (E-F) +j\epsilon} \right) , \label{chi0}
\end{eqnarray}
\noindent where the sum runs over transitions between filled and empty orbitals
with energies $E$ and $F$, $j^2=-1$, and $\epsilon$ is a regularisation
parameter. Since the orbitals $\psi_E$ may be expressed as linear combinations
of atomic orbitals (AO's), this form of $\chi_0$ exhibits dependence on AO pair
products.

References~\cite{Foerster2008,Foerster2009} point out the high degree of linear
dependence in the AO product space and the means to drastically reduce it by
expressing AO products as linear combinations of dominant products found by
diagonalisation of an appropriate metric. Paper~\cite{Koval2010} avoids
explicit inversion in eqn.~(\ref{chitot}) and introduces an efficient parallel
solution of the relevant equations:

\begin{equation}
\left \{ 
\begin{array}{l}
\displaystyle P(\omega) = \sum_{i=1}^{3}{<d_i,X_i(\omega)>} \\
\\
\left(1 - \chi^{0}(\omega)\Sigma\right)X_i(\omega) =\chi^{0}(\omega) d_i, \qquad i=1,2,3
\end{array}
\right. \label{grmreseqs}
\end{equation} 
where $d_i$ is the dipole in the $i$-direction. Each linear system is solved by
the Krylov GMRES method\cite{Saad2003}.  Solution of eqns. (\ref{grmreseqs}) at
a set of frequencies $\omega$  leads to a raw absorption
spectrum.

\begin{figure}
\begin{center}
\includegraphics[scale=.6, angle=0]{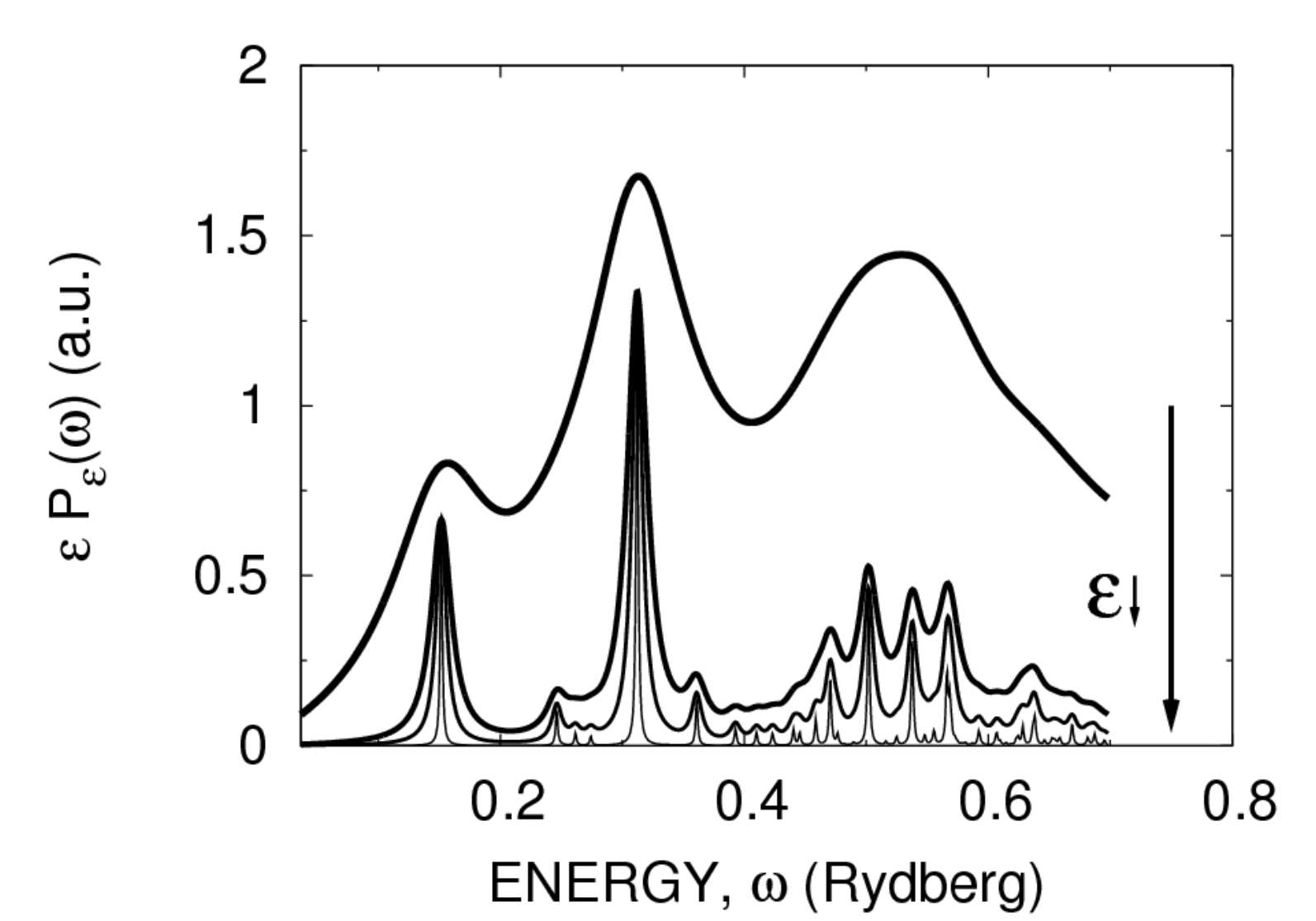} 
\end{center}
\caption{\label{epsdep}Absorption spectra of indigo computed with regularly
spaced frequency points and decreasing regularisation parameter $\epsilon$=0.05, 0.01, 0.005 and 0.001 Rydberg,
from top to bottom.}
\end{figure}

Note that because $\epsilon$ is a regularisation parameter,  the raw spectrum
for a particular value of $\epsilon$ has no absolute physical meaning.  Figure
(\ref{epsdep}) shows how, given sufficient points in the raw spectrum, close
resonances are distinguishable as the regularisation parameter is reduced.  For
convenience of representation, we plot $\epsilon P_{\epsilon}(\omega)$ rather
than $P_{\epsilon}(\omega)$, which diverges at resonances as $\epsilon
\rightarrow 0$. Finding with any certainty even just the main resonances in a
given frequency interval $\left[ \omega_{min},\omega_{max}\right]$ would seem
to require making $\epsilon$ very small and using a very large number of
points, of order $(\omega_{max} - \omega_{min})/\epsilon$. However, quite apart
from the computational cost, it is ineffective to try to separate close
resonances by brute force reduction of $\epsilon$ and increasing the number of
frequency points $\omega$. Close to resonances of the free electron response,
$\omega \sim E-F$  in eqn. (\ref{chi0}), $\chi_0$ diverges as $1/\epsilon$
and ill-conditioning may prevent convergence of the {\sc gmres} method. We have
observed slow convergence, or absence of convergence, or even negative
polarisabilities at some frequency points when $\epsilon$ is reduced below
$5\times10^{-4}$\,Ry. There is no clear way to preconditioning the linear
systems (\ref{grmreseqs}), so some other strategy is needed. 

Furthermore, the method as it stands has other drawbacks:  {\em (i)} The shape
of the spectrum depends on the regularisation parameter $\epsilon$. When it is
too big, a weak transition may go unnoticed in the wing of a strong one; but
making it too small is wasteful, with most values of $P(\omega)$ very small
compared to a handful of large values at the resonances; {\em (ii)} Little can
be done to identify the nature of the transitions observed in the spectrum,
since not even oscillator strengths are extracted; {\em (iii)} The \tddft
computation is run after \siesta, requiring geometrical, orbital, Hamiltonian
and overlap data to be communicated in a disk file.

\subsubsection*{Present improvements}
An immediate step to improving separation of resonances  is to deal separately
with each polarisation  in eqn.  (\ref{grmreseqs}), since transitions often have
different polarisations.  Here we further improved the computation in several
ways.

First, addressing point {\em(iii)}, the '\fast' \tddft calculation can be invoked
now from within \siesta, the 'move' loop of the \siesta\ main program. This is
achieved by coupling \fast\ directly to \siesta\ using the \mpicpl\ (MPI Coupling) framework\cite{mpicpl}.
\mpicpl\ is dedicated to the coupling of scientific codes, based on the
well-known MPI standard. It is divided into several independent layers for
coupling, data redistribution and steering. The codes to be coupled are
launched and connections between them are set up by mpicpl, according to
information derived from an xml file.

Second, to address points  {\em (i)} and {\em (ii)}, note that the expected form of the
spectrum is a sum of Lorentzian resonances. By fitting the parameters of these
Lorentzians to the numerical spectrum, we can in most cases identify the
transitions without recourse to small $\epsilon$ and very fine combs of
$\omega$ values.   We relate the peak heights of the numerical spectrum to the
oscillator strengths of the transitions by considering that in the linear
response regime, electronic transitions respond to  the driving field like
independent harmonic oscillators~\cite{Mukamel1995}, so that the susceptibility
$$\chi(\omega) = \chi'(\omega) + j \chi''(\omega) $$ can be written in atomic
units as
\begin{equation*}
\chi(\omega)=\sum_I \frac{f_I}{\Omega_I^2-\omega^2-2j\omega\Gamma_I} ,
\end{equation*}

\noindent where $\Omega_I$, $f_I$ and $\Gamma_I$ are the frequency, oscillator
strength and damping (homogeneous linewidth) of transition $I$. Here, we need
the complex part of the susceptibility, $\chi''(\omega)$, which after a little
algebra can be expressed as a sum of terms of the form
$$
\frac{2\omega\Gamma_I f_I}{(\Omega_I - \omega)^2(\Omega_I + \omega)^2 + 4\omega^2 \Gamma_I^2}.
$$
\noindent Close to a resonance (elsewhere the susceptibility is negligible) ,
$\omega \sim \Omega_I$, and $\Omega_I + \omega \sim 2\omega \sim
2\Omega_I $, so that 
\begin{equation*}
  \chi''(\omega) \sim \sum_I   \frac{f_I \Gamma_I/2\Omega_I}{(\Omega_I - \omega)^2 + \Gamma_I^2}.
\end{equation*}

\noindent Identifying the regularisation parameter $\epsilon$ in the linear
response  \tddft\ with the damping $\Gamma_I$ , we see that the numerical
spectrum should be representable as a sum of normalised (unit area) Lorentzian
resonances,
\begin{equation}
\chi''_{\tsub{model}}(\omega) \sim \sum_{I=1}^{N_{\tsub{res}}} \frac{C_I \epsilon / \pi}{(\Omega_I - \omega)^2 + \epsilon^2} + B,
\label{modelspec}
\end{equation}

\noindent where the weights $C_I$ are related to the oscillator strengths by
\begin{equation}
\frac{f_I}{2 \Omega_I} = \frac{C_I}{\pi} . \nonumber
\end{equation}
\noindent In eqn. (\ref{modelspec}), $B$ represents the more or less flat contribution
of resonances outside the frequency range $\left[
\omega_{\tsub{min}},\omega_{\tsub{max}}\right]$ where the raw spectrum was
computed. We perform a non-linear least squares fit of the parameters $C_I$, $\Omega_I$ and $B$ to minimise the
residual:

\begin{equation}
\chi^2 = \frac{1}{N_{\tsub{freq}}} \sum_{j=1}^{N_ {\tsub{freq}}} \left( \chi''_{\tsub{model}}(\omega_j) - I_j \right)^2 , \label{residualfit}
\end{equation}
%

\noindent using  the Levenberg-Marquardt method from the {\sc gnu} scientific
library~\cite{GSL} and $N_{\tsub{freq}}$ frequency points  $\omega_j$ in the
numerical \tddft\ spectrum, with intensities $I_j$. 

Fitting a Lorentzian requires at least four data points. If $N_L$ resonances
are expected in the  frequency range
$[\omega_{\tsub{min}},\omega_{\tsub{max}}]$, the raw spectrum should
comprise at least $N_{\tsub{freq}}=N_1=4N_L$ points. A reasonable trial
$\epsilon$ would be $\epsilon=\epsilon_1=\left( \omega_{\tsub{max}} -
\omega_{\tsub{min}}\right)/N_1$. The number of certain resonances and their
positions are determined by an inspection algorithm detecting local maxima of
the spectrum from adjacent regions of increasing or decreasing values.
Clearly, the smaller the value of $\epsilon$, the sharper the resonances will
become, and the greater will be the number of distinguishable resonances. But
because of the {\em caveat}\ noted above and of the computational cost,
$\epsilon$ should not be made too small.  

The question now is how to best choose the $\omega_j$. In practice, data in the
wings of the resonances contribute less to the accuracy of the Lorentzian fit
than do points close to the peaks.  Since the resonances are at first unknown,
one must start with a uniform distribution of $\omega_j$ but it is possible to
continue with an iterative, adaptive procedure, to cluster the points  around
the current best approximations to the
resonance frequencies.  
\begin{figure}
\begin{center}
\includegraphics[scale=.5]{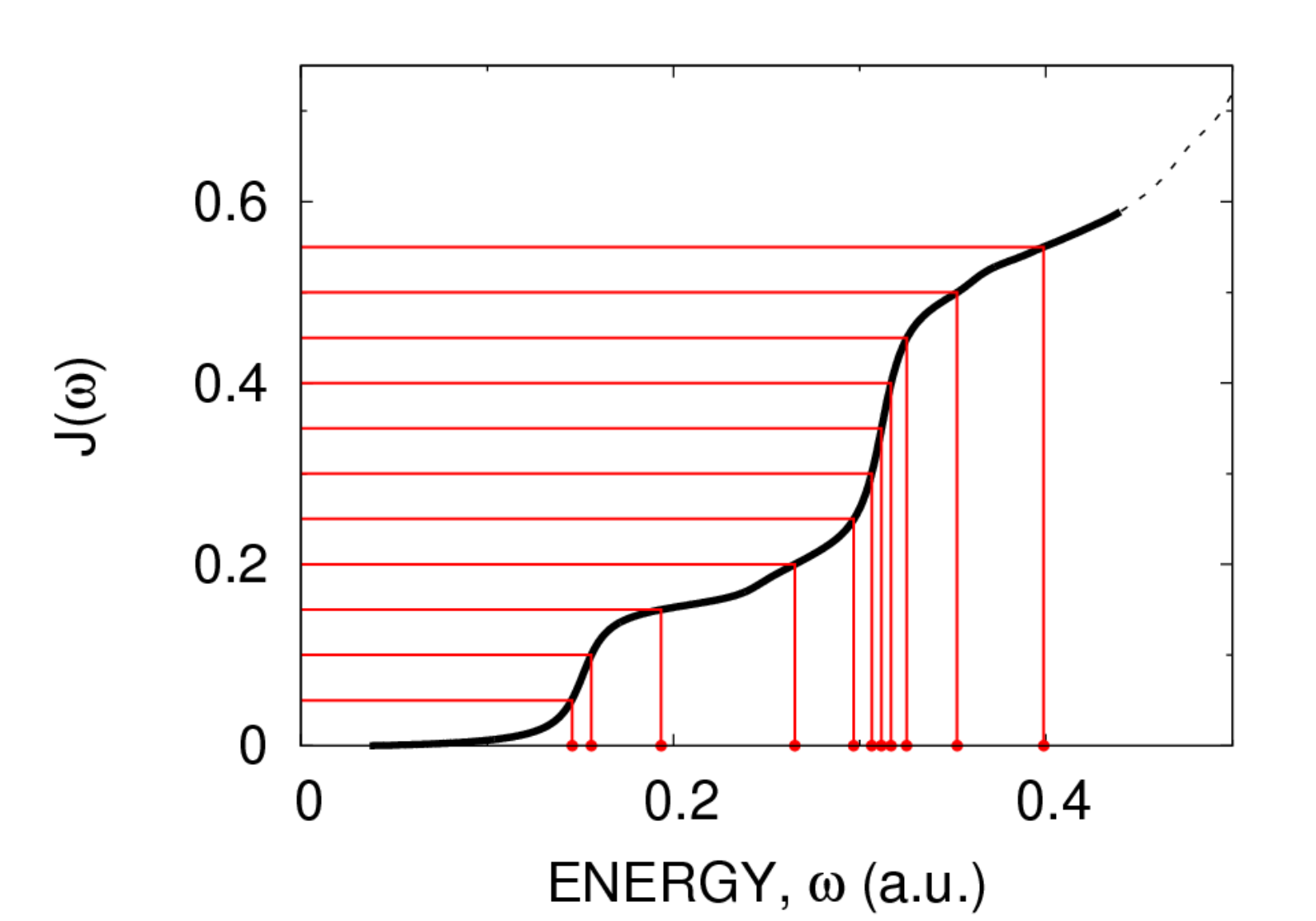} 
\end{center}
\caption{ Behaviour of the integral function $J(\omega)$ used to to concentrate
\tddft\ computation points  $\omega_j$  around the resonances of the current
best estimate of the spectrum (in this case the top spectrum in figure
\ref{epsdep}).  \label{stepup}}
\end{figure}

With this purpose, define a normalized integral function as
\begin{equation}
J(\omega)= \left. \displaystyle\int_{\omega_{\tsub{min}}}^\omega P(\omega)d\omega \middle/ \displaystyle\int_{\omega_{\tsub{min}}}^{\omega_{\tsub{max}}} P(\omega)d\omega \right. .\label{integral}
\end{equation}

As $P$ is a positive function, $J$ increases monotonically (from 0 to 1) as
$\omega$ sweeps the interval $[\omega_{\tsub{min}},\omega_{\tsub{max}}]$. In
practice, $J(\omega)$ increases step-wise at each resonance, the smaller the
value of $\epsilon$, the steeper the steps and the flatter the plateaux between
the steps, see fig. (\ref{stepup}). Therefore, we can easily construct a local
inverse function where $P$ is non zero. The trick is now to use the inverse
function to map a set of regularly spaced values of $J$ into a set $\omega$'s
clustered around the resonances. Since at any time we know  only a finite
number of $(\omega_i,J(\omega_i))$ pairs, the local inverse function is defined
{\em via} linear interpolation:  
\begin{equation}
J_{\hbox{\scriptsize approx}}(\omega)=J(\omega_i)\frac{J(\omega_{i+1})-J(\omega_{i})}{\omega_{i+1}-\omega_{i}}(\omega-\omega_i)\qquad \omega\in[\omega_{i},\omega_{i+1}].\label{discreteintegral}
\end{equation}
\noindent Figure (\ref{stepup}) illustrates this procedure for the topmost
spectrum ($\epsilon=0.05$\,a.u.) in figure (\ref{epsdep}).

The complete algorithm is provided in the supplementary
information, with results for uniformly spaced data.
Briefly, we start with a uniform distribution of points and
iterate the clustering around the current best estimates of the resonances.  At
each stage of the procedure, the number of certain resonances to be used in the
fit is determined by inspection of the numerical spectrum, and the inverse
integral function is used to improve the distribution of the points around
these resonances before refitting the Lorentzians. After iterating the
improvement of the distribution of points, the algorithm may decrease
$\epsilon$ or increase the number of points, or do both. 

\subsubsection*{Application to indigo}

\begin{figure}
\begin{minipage}[t][1\totalheight]{0.7\textwidth}
\begin{center}
\begin{minipage}[t][1\totalheight]{0.3\textwidth}
\includegraphics[scale=.3, angle=0]{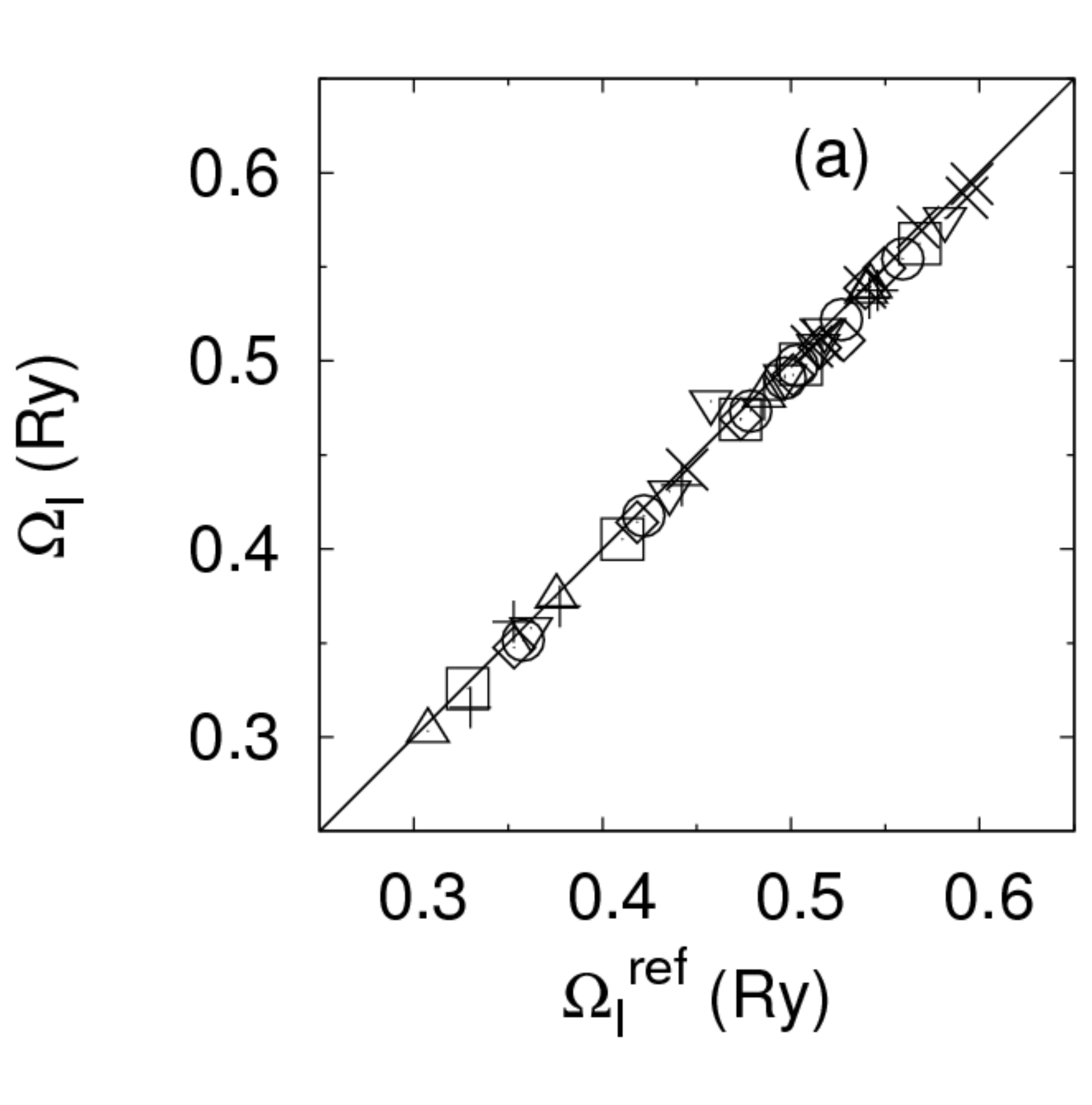} 
\end{minipage}
\hfill
\begin{minipage}[t][1\totalheight]{0.3\textwidth}
 \includegraphics[scale=.3, angle=0]{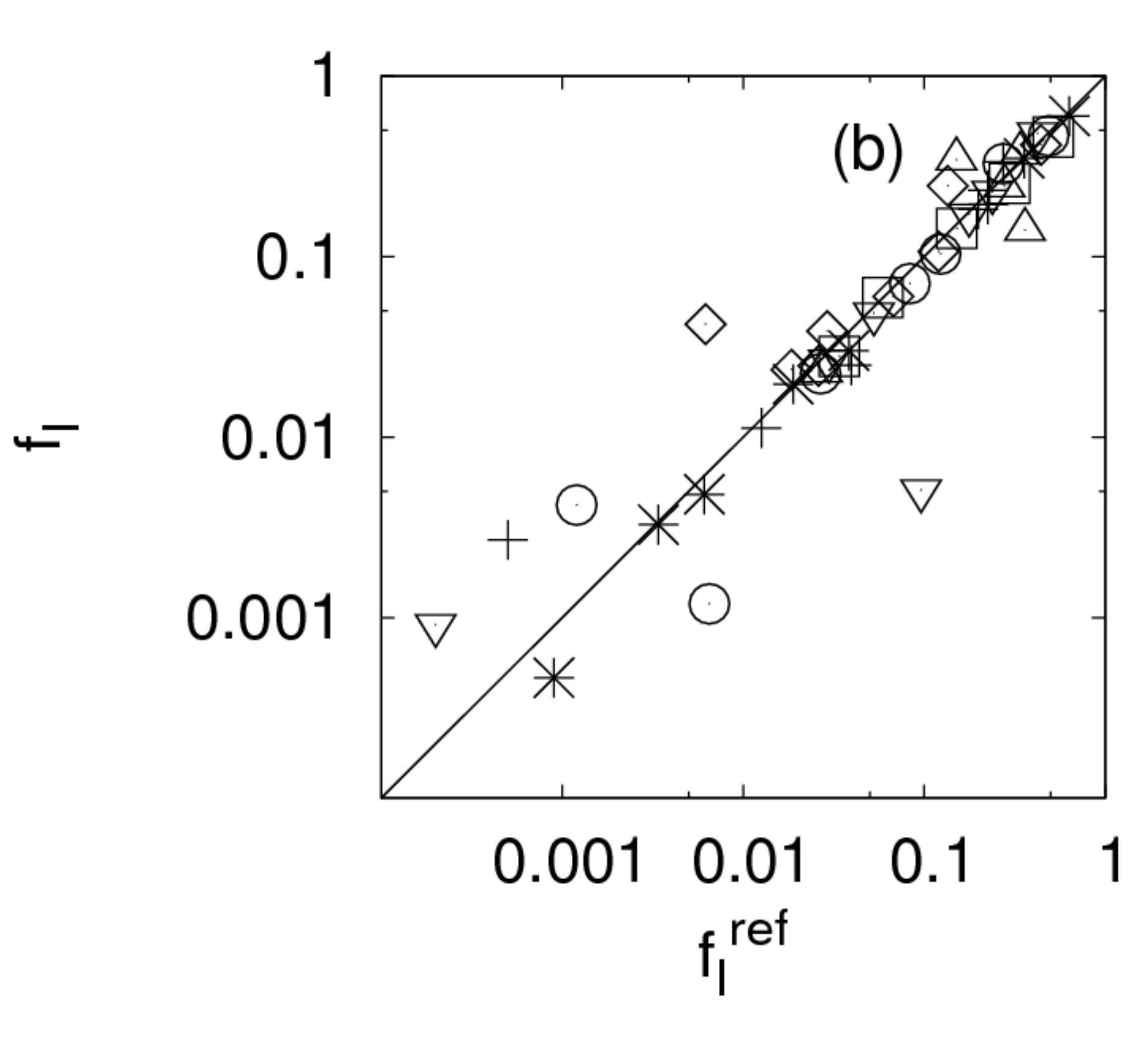}
\end{minipage}
\end{center}
\end{minipage}
\caption{Comparison of transitions obtained with the present adaptive algorithm (LDA, numerical DZP basis, vertical axis) to those obtained by solution of Casida's eigenproblem equations ( LSD, 6-31G*, horizontal axis), for a series of substituted benzenes. {\em (a)}\ Transition energies;  {\em (b)}\ oscillator strengths. Symbols : $\Box$ aniline; $\circ$ anisole; $\bigtriangleup$ dimethylaniline; $\bigtriangledown$ phenol; $\diamond$ phentole; $+$ thiophenol; $\times$ toluene .\label{comparetrans}} 
\end{figure}

Before going into more detail about the choice of adaptive strategies in the algorithm, figure (\ref{comparetrans}) validates the present approach. It shows results for a set of substituted benzenes. The figure compares transition energies and oscillator strengths obtained here (\lda, standard {\sc dzp} valence basis and pseudo-potentials in \siesta) to those obtained with a similar level of theory ({\sc lsd}, 6-31G* all electron basis) by solving Casida's eigenproblem casting of linear response theory\cite{Casida1995,Jamorski1996,gaussian09}. Calculations were performed on the same geometries, optimised in \siesta. Despite differences in the basis sets, the agreement is good, including even most of the weaker transitions.

\begin{table}[htbp]
\centering\begin{tabular}{|l|c|c|c|c|c|}
\hline
      & Test 1 & test 2& test 3& test 4& test 5 \\ \hline
$\epsilon^{\tsub{initial}}$ (Ry)      & $1 \times 10^{-3}$ & $1 \times 10^{-3}$  & $1 \times 10^{-3}$ & $1 \times 10^{-3}$ & $2 \times 10^{-2}$\\
$\epsilon^{\tsub{final}}$ (Ry)        & $1 \times 10^{-3}$ & $1 \times 10^{-3}$  & $1 \times 10^{-3}$ & $1 \times 10^{-3}$ & $1 \times 10^{-3}$\\
                                      &                    &                     &                    &     &   \\
$N_{\tsub{freq}}^{\tsub{initial}}$    & 571                & 285                 & 101                & 101 & 50\\ 
$N_{\tsub{freq}}^{\tsub{final}}$      & 571 & 285 & 101 & 198 & 255\\ 
                                      &     &     &     &     &     \\
$N_{\tsub{iter}}$        & 2   & 2   &  5  & 4   &  5\\
$N_{\tsub{eval}}$   & 1142& 570 & 505 & 583 & 663\\
                    &     &     &     &     &    \\
$N_{\tsub{trans}}$ & 7   & 7   &  4  &   6 &  7\\
$\chi/||I||_2$   &  $3.0 \times 10^{-4}$  & $3.4 \times 10^{-4}$& $1.7 \times 10^{-3}$&$5.9 \times 10^{-3}$& $4.2 \times 10^{-4}$ \\
$\delta f_{\tsub{max}}$        &$ 2.2\times 10^{-4}$&$2.1\times 10^{-5}$&$1.0 \times 10^{-4}$&  $8.9 \times 10^{-4}$ &$6.2 \times 10^{-4}$\\ 
\hline 
\end{tabular}

\caption{Statistics of different iterative strategies to improve the fitting of Lorentzians to the numerical \tddft\ spectrum of indigo. $N_{\tsub{freq}}^{\tsub{initial}}$, $N_{\tsub{freq}}^{\tsub{final}}$ : initial and final numbers of data points; $N_{\tsub{iter}}$ : number of iterations to reach an error of $5.0 \times10^{-2}$\,Ry ; $N_{\tsub{eval}}$ : total number of evaluations of  $P(\omega)$ at convergence;  $N_{\tsub{trans}}$ : number of transitions found;  $\chi^2$ : 
the residual defined by eqn. (\ref{residualfit}). $\delta f_{\tsub{max}}$ is the maximum variation of the oscillator strengths between the last two iterations at convergence.
\label{Tab:Adap1}}

\end{table}
Table~\ref{Tab:Adap1} illustrates four ways to use the adaptive fitting to
improve the accuracy on the transitions obtained for the visible spectrum of
indigo in the range 0.02 to 0.4\,Ry, computed with the same \dft\ conditions as
in section \ref{ESP}. The final value of the regularization parameter is in
each case $\epsilon=10^{-3}$\,Ry. Table~\ref{Tab:Adap1} shows two strategies.
Cases 1--4 illustrate constant $\epsilon$ ($10^{-3}$\,Ry) and variable placing
or numbers of data points. In case 5, both  $\epsilon$ and the number of data
points are varied, the latter being determined by $\epsilon$. The idea, here is
to use fewer points placed depending on the regularized parameter.  
We summarize below the parameters of the different strategies :
\begin{itemize}
\item[]\textit{Test 1} 
Here,  the number of points is constant and we iteratively improve the point
placement. Considering that resonances will not be separable if closer than
$\epsilon$, up to 
\begin{equation}
 N_{\tsub{trans}} \sim (\omega_{\tsub{max}}-\omega_{\tsub{max}})/2\epsilon 
 \label{nres}
\end{equation}
\noindent resonances could in principle be distinguished, given 4 points for
each. The minimum number of points required is thus:
\begin{equation}
N_{\tsub{freq}} = 1 + \frac{3}{2\epsilon}( \omega_{max}- \omega_{min}). \label{eqn:automaticN}
\end{equation}
\noindent  

\item[]\textit{Test 2} and \textit{test 3} are the same as test 1, but use
fewer points, respectively 285 and 101.

\item[]\textit{Test 4} starts with the same small number of points as
\textit{test3}, and at each iteration $N_{\tsub{freq}}$ is increased by $25$\%. 

\item[]\textit{Test 5} starts with $\epsilon = 2 \times 10^{-2}$\,Ry and $N_{\tsub{freq}} =50$ 
and $\epsilon$ is reduced in equal steps to $10^{-3}$\,Ry in two iterations.  At each iteration the number of points increases by 25\,\% in order to have enough points for the fit algorithm. 

\end{itemize}
Table~\ref{Tab:Adap1} shows the consequences of these strategies. \textit{Test 1}\
represents a brute force approach. But most real molecules will have far fewer
transitions than implied by eqn (\ref{nres}). Accordingly, \textit{tests 2--3} show that
the number of points can be reduced without loss of information, but that
eventually (\textit{test 3}) some weak transitions are missed. These are recovered in
\textit{test 4}, where data points are added, but only as necessary. The number of
evaluations of $P(\omega)$ is half that required by brute force, for a
comparable result. \textit{Test 5} achieves an even better result, at the expense of
slightly more function evaluations, by decreasing $\epsilon$.

\begin{figure}
\begin{center}
\includegraphics[scale=.3, angle=0]{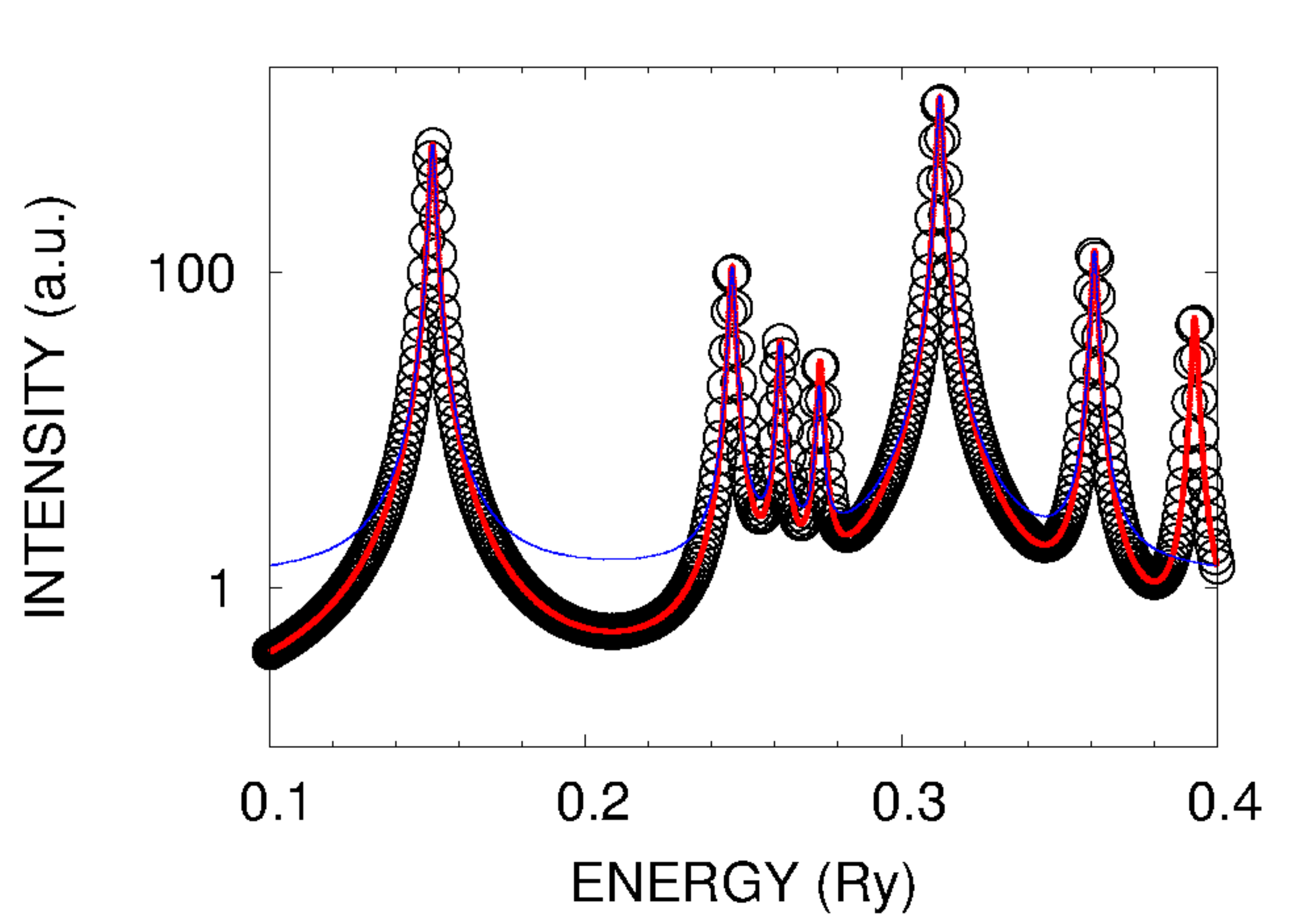} 
\caption{ The UV-visible part of the absorption spectrum of indigo; semi-log scale to highlight strong and weak resonances. Data points : numerical spectrum; thin  line : the Lorentzian fit of test 4 misses the highest energy resonance, which is recovered in test 5 (thick line).}\label{Fig:adaptif}
\end{center}
\end{figure}

Figure~\ref{Fig:adaptif} shows the spectrum and the fits obtained in \textit{tests  4} and \textit{5}. Resonances found in the tests agree very well, to within about $10^{-5}$\,Ry (4\wave) for the resonance frequency and about 0.1\,\% for the oscillator strengths. These results illustrate a further benefit of the fitting procedure, that, using a moderate $\epsilon$ ($10^{-3}$\,Ry),  it achieves an accuracy that could only be achieved by brute force with much smaller $\epsilon$, for which the \tddft\ algorithm would be numerically unstable. The energy and oscillator strength of the first excited state of indigo are  0.1515\, Ry (or a wavelength of 602\,nm in vacuum) and 0.2, values which could certainly be improved (the experimental transition is at 0.136\,Ry or 546\,nm\cite{Sadler1956}), should hybrid functionals become available in \siesta. 

\section{Concluding remarks\label{conclusion}}
We have described several extensions of the widely used \siesta\ program, aimed
at making it more directly applicable to molecular systems. Some closing words may be
appropriate on the methods chosen.

Self polarisation in the periodic model of an isolated water molecule was
small. However, larger effects can be expected for systems which are both polar
and contain $\pi$ electrons, such as many common laser dyes, in which a degree
of intramolecular charge transfer is present, even in the ground state. Large
dipole moments are common in laser dyes, which moreover may contain 20--40
heavy atoms. Simulation boxes big enough to effectively separate such systems
from their images could become uncomfortably large. The present multi-grid
solver for the Poisson equation with Dirichlet boundary conditions meets this need, in a framework that should make further improvements possible, such as use of finite elements to represent the usually inhomogeneous electronic density.  

Electrostatic potential fitting is an old problem, with a vast literature. The
algorithm introduced here in \siesta\ is very simple. While it so far has
worked satisfactorily for molecules of interest to us, general users should be
aware that it may produce unpredictable results on systems with 'buried' atoms.
The algorithm here has also been adapted to fit the electrostatic potential
computed in \siesta\ in periodic systems.  See however Campa\~na
\etal\cite{Campana2009} for a lucid discussion of the problems associated with
charge fitting in such systems.

The design atom approach to truncation of molecules does not appear to have
been taken up in the literature, possibly because Xiao and Zhang exhibited
significant perturbation of the truncated moieties of {\em small}
molecules\cite{Xiao2007}. Indeed, we too found for example that dihedrals close
to the design atom may be in error by 10--20\,\dang. Yet, as shown here, the
perturbation in fact decays rapidly with distance from the design atom, making
the method attractive for truncation of large systems.

It should also be mentioned that the lone pair of the design atom may give rise to
spurious $n-\pi*$ transitions towards a conjugated region of the truncated
molecule. They appear as weak transitions in the low energy wing of
the spectrum, with intensities falling off exponentially with the distance to the
design atom. This easily identified effect is tolerable in large molecules, in
view of the gain in computational cost brought by truncation. A typical case
would be truncation in quantum mechanical/classical mechanical simulations,
where a number of other approximations might be be more serious than the
presence of these weak but identified transitions

It may be useful to put the linear-response \tddft\ method in refs.
\cite{Foerster2008,Foerster2009,Koval2010} and the present improvements in
perspective with commoner approaches, such as for example Casida's widely
implemented equations\cite{Casida1995,Jamorski1996}. The point is that whereas solution of Casida's
equations yields at significant cost a list of transitions and their
properties, the linear response \tddft\ method produces a piece of the raw
spectrum, at low cost, but with little physical insight, such as orbital
symmetries and energies, transition polarisations and oscillator strengths. 

The present extensions, by  efficiently extracting transition energies,
oscillator strengths and a degree of polarisation information, should be useful
to identify transitions by comparison with a one-off solution of Casida's
equations for the same system. The strength of the present method would then be to
allow cheap, repetitive calculations to study how the transition reacts to
multiple perturbations of the geometry, or a varying external field, because
$P(\omega)$ then needs to be computed only in narrow frequency intervals
bracketing the interesting resonances.  Indeed, useful experimental data are in the vast
majority of cases restricted to the first one or two strongest excited states only. For
example, the first transition of indigo on a recent 12-core computer could be
computed in less 30\,s\, of which over 90\%\ of time was spent building the Hartree
and exchange-correlation contributions (half each) to the interaction kernel.
The cost of the iterative procedure was negligible. Computing spectra on the fly during molecular dynamics, \eg\  solvation
shifts in liquids, is thus a realistic prospect of the present methods.

\vspace{10mm}
\begin{acknowledgements}
It is pleasure to thank members of the \siesta\ development team, particularly
Alberto Garc\'ia, {\sc icmab}, Barcelona, for their encouragement and advice.
Part of the calculations were performed at the high performance computing
facility MCIA: M\'esocentre de calcul intensif en Aquitaine and on the PlaFRIM
experimental testbed, developed under the Inria PlaFRIM development action with
support from LABRI and IMB and other entities: Conseil R\'egional d'Aquitaine,
FeDER, Universit\'e de Bordeaux and CNRS (see
https://plafrim.bordeaux.inria.fr/).
\end{acknowledgements}

\bibliographystyle{spmpsci}      

\bibliography{nossi}   

\begin{thebibliography}{10}
\providecommand{\url}[1]{{#1}}
\providecommand{\urlprefix}{URL }
\expandafter\ifx\csname urlstyle\endcsname\relax
  \providecommand{\doi}[1]{DOI~\discretionary{}{}{}#1}\else
  \providecommand{\doi}{DOI~\discretionary{}{}{}\begingroup
  \urlstyle{rm}\Url}\fi

\bibitem{GSL}
{GSL} - {GNU} {S}cientific {L}ibrary.
\newblock Http://www.gnu.org/software/gsl

\bibitem{hypre}
hypre: High performance preconditioners.
\newblock {h}ttp://www.llnl.gov/CASC/hypre

\bibitem{mpicpl}
{MPICPL}: {MPI} coupling.
\newblock {h}ttps://gforge.inria.fr/projects/mpicpl/

\bibitem{hypreManual}
{hypre} User's manual, version 2.0.0 (2006).
\newblock {Center for Applied Scientific Computing, Lawrence Livermore National
  Laboratory, Livermore, CA 94550}

\bibitem{lapack}
Anderson, E., Bai, Z., Bischof, C., Blackford, S., Demmel, J., Dongarra, J.,
  Du~Croz, J., Greenbaum, A., Hammarling, S., McKenney, A., Sorensen, D.:
  {LAPACK} Users' Guide, third edn.
\newblock Society for Industrial and Applied Mathematics, Philadelphia, PA
  (1999)

\bibitem{Artacho2008}
Artacho, E., Anglada, E., Dieguez, O., Gale, J.D., Garc\'ia, A., Junquera, J.,
  Martin, R.M., Ordej\'on, P., Pruneda, J.M., S\'anchez-Portal, D., Soler,
  J.M.: The siesta method; developments and applicability.
\newblock J. Phys.: Condens. Matter \textbf{20}, {064,208--1}--{6} (2008)

\bibitem{Bayly1993}
Bayly, C.I., Cieplak, P., Cornell, W.D., Kollman, P.A.: A well-behaved
  electrostatic potential based method using charge restraints for deriving
  atomic charges: the resp method.
\newblock J. Phys. Chem. \textbf{97}, 10,269--10,280 (1993)

\bibitem{Briggs2000}
Briggs, W.L., Henson, V.E., McCormick, S.F.: A Multigrid Tutorial.
\newblock Siam, Philadelphia, PA (2000)

\bibitem{Campana2009}
{Campa\~{n}a}, C., Mussard, B., Woo, T.K.: Electrostatic potential derived
  atomic charges for periodic systems using a modified error functional.
\newblock J. Chem. Theory Comput. \textbf{5}, 2866--2878 (2009)

\bibitem{Casida1995}
Casida, M.E.: Time dependent density functional response theory for molecules.
\newblock In: D.P. Dong (ed.) Recent advances in density functional methods,
  part I, Recent advances in computational chemistry, pp. 155--192. World
  Scientific, Singapore (1995)

\bibitem{Connolly1983}
Connolly, M.L.: {Analytical molecular surface calculation}.
\newblock Journal of Applied Crystallography \textbf{16}(5), 548--558 (1983)

\bibitem{hypreSolver}
Falgout, R.D., Jones, J.E.: Multigrid on massively parallel architectures.
\newblock In: E.~Dick, K.~Riemslagh, J.~Vierendeels (eds.) Multigrid Methods
  VI, \emph{Lecture Notes in Computational Science and Engineering}, vol.~14,
  pp. 101--107. Springer, Berlin (2000).
\newblock Proc. of the Sixth European Multigrid Conference held in Gent,
  Belgium, September 27-30, 1999. Also available as LLNL technical report
  UCRL-JC-133948

\bibitem{hypreDesign}
Falgout, R.D., Jones, J.E., Yang, U.M.: The design and implementation of {\sl
  hypre}, a library of parallel high performance preconditioners.
\newblock In: A.M. Bruaset, A.~Tveito (eds.) Numerical Solution of Partial
  Differential Equations on Parallel Computers, pp. 267--294. Springer--Verlag
  (2006).
\newblock Also available as LLNL technical report UCRL-JRNL-205459

\bibitem{Foerster2008}
Foerster, D.: Elimination, in electronic structure calculations, of redundant
  orbital products.
\newblock J. Chem. Phys. \textbf{128}, {034,108--1}--{9} (2008)

\bibitem{Foerster2009}
Foerster, D., Koval, P.: On the {Kohn-Sham} density response in a localized
  basis set.
\newblock J. Chem. Phys. \textbf{131}, {044,103--1}--{9} (2009)

\bibitem{MillerFrancl1996}
Francl, M.M., Carey, C., Chirlian, L.E.: Charges fit to electrostatic
  potentials. ii. {C}an atomic charges be unambiguously fit to electrostatic
  potentials.
\newblock J. Comp. Chem. \textbf{17}, 367--383 (1996)

\bibitem{gaussian09}
Frisch, M.J., Trucks, G.W., Schlegel, H.B., Scuseria, G.E., Robb, M.A.,
  Cheeseman, J.R., Scalmani, G., Barone, V., Mennucci, B., Petersson, G.A.,
  Nakatsuji, H., Caricato, M., Li, X., Hratchian, H.P., Izmaylov, A.F., Bloino,
  J., Zheng, G., Sonnenberg, J.L., Hada, M., Ehara, M., Toyota, K., Fukuda, R.,
  Hasegawa, J., Ishida, M., Nakajima, T., Honda, Y., Kitao, O., Nakai, H.,
  Vreven, T., Montgomery {Jr.}, J.A., Peralta, J.E., Ogliaro, F., Bearpark, M.,
  Heyd, J.J., Brothers, E., Kudin, K.N., Staroverov, V.N., Kobayashi, R.,
  Normand, J., Raghavachari, K., Rendell, A., Burant, J.C., Iyengar, S.S.,
  Tomasi, J., Cossi, M., Rega, N., Millam, J.M., Klene, M., Knox, J.E., Cross,
  J.B., Bakken, V., Adamo, C., Jaramillo, J., Gomperts, R., Stratmann, R.E.,
  Yazyev, O., Austin, A.J., Cammi, R., Pomelli, C., Ochterski, J.W., Martin,
  R.L., Morokuma, K., Zakrzewski, V.G., Voth, G.A., Salvador, P., Dannenberg,
  J.J., Dapprich, S., Daniels, A.D., Farkas, Â., Foresman, J.B., Ortiz, J.V.,
  Cioslowski, J., Fox, D.J.: Gaussian~09 {R}evision {A}.1.
\newblock Gaussian Inc. Wallingford CT 2009

\bibitem{Greengard1987}
Greengard, L., Rokhlin, V.: A fast algorithm for particle simulations.
\newblock Journal of Computational Physics \textbf{73}(2), 325 -- 348 (1987)

\bibitem{Jackson1998}
Jackson, J.D.: {Classical Electrodynamics}, third edn.
\newblock Wiley (1998)

\bibitem{Jamorski1996}
Jamorski, C., Casida, M.E., Salahub, D.R.: Dynamic polarizabilities and
  excitation spectra from a molecular implementation of time-dependent
  density-functional response theory: N$_2$ as a case study.
\newblock J. Chem. Phys. \textbf{104}, 5134--5147 (1996)

\bibitem{Koval2010}
Koval, P., Foerster, D., Coulaud, O.: A parallel iterative method for computing
  molecular absorption spectra.
\newblock J. Chem. Theory Comput. \textbf{6}, 2654--2668 (2010)

\bibitem{Lacombe2009}
Lacombe, S., Soumillion, J.P., El~Kadib, A., Pigot, T., Blanc, S., Brown, R.,
  Oliveros, E., Cantau, C., Saint-Cricq, P.: Solvent-free production of singlet
  oxygen at the gas-solid interface: Visible light activated organic-inorganic
  hybrid microreactors including new cyanoaromatic photosensitizers.
\newblock Langmuir \textbf{25}(18), 11,168--11,179 (2009)

\bibitem{blas}
Lawson, C.L., Hanson, R.J., Kincaid, D.R., Krogh, F.T.: Basic linear algebra
  subprograms for fortran usage.
\newblock ACM Trans. Math. Softw. \textbf{5}(3), 308--323 (1979)

\bibitem{Mukamel1995}
Mukamel, S.: Principles of nonlinear optical spectroscopy.
\newblock O.U.P., New York (1995)

\bibitem{Perdew1981}
Perdew, J.P., Zunger, A.: Self-interaction correction to density-functional
  approximations for many-electron systems.
\newblock Phys. Rev. B \textbf{23}, 5048--5079 (1981)

\bibitem{Saad2003}
Saad, Y.: Iterative Methods for Sparse Linear Systems.
\newblock Siam, Philadelphia (2003)

\bibitem{Sadler1956}
Sadler, P.W.: Absorption spectra of indigoid dyes.
\newblock J.Org. Chem. \textbf{21}, 316--318 (1956)

\bibitem{Sanz-Navarro2011}
Sanz-Navarro, C.F., Grima, R., Garc\'ia, A., Bea, E.A., Soba, A., Cela, J.M.,
  Ordej\'on, P.: An efficient implementation of a {\sc qm-mm} method in {\sc
  siesta}.
\newblock Theor. Chem. Acc. \textbf{128}, 825--833 (2011)

\bibitem{Sigfridsson1998}
Sigfridsson, E., Ryde, U.: Comparison of methods for deriving atomic charges
  from the electrostatic potential and moments.
\newblock J. Comp. Chem. \textbf{19}, 377--395 (1998)

\bibitem{Soler2002}
Soler, J.M., Artacho, E., Gale, J.D., Garc\'ia, A., Junquera, J., Ordej\'on,
  P., S\'anchez-Portal, D.: The {S}iesta method for ab initio order-n materials
  simulation.
\newblock J. Phys.: Condens. Matter \textbf{14}, 2745--2779 (2002)

\bibitem{Terrabuio2012}
Terrabuio, L.A., Haiduke, R.L.A.: Electrostatic potentials and polarization
  effects in proton-molecule interactions by means of multipoles from the
  quantum theory of atoms in molecules.
\newblock International Journal of Quantum Chemistry \textbf{112}(19)

\bibitem{Xiao2007}
Xiao, C., Zhang, Y.: Design-atom approach for the quantum mechanical/molecular
  mechanical covalent boundary: A design-carbon atom with five valence
  electrons.
\newblock J. Chem. Phys. \textbf{127}, {124,102--1}--{9} (2007)

\end{thebibliography}


\newpage
\centerline{\LARGE\bf Supplementary Information}
\setcounter{section}{0}

\section{Linear scaling of the MG solver of the Poisson equation}
\begin{figure}[h]
\begin{center}
\includegraphics[scale=.4]{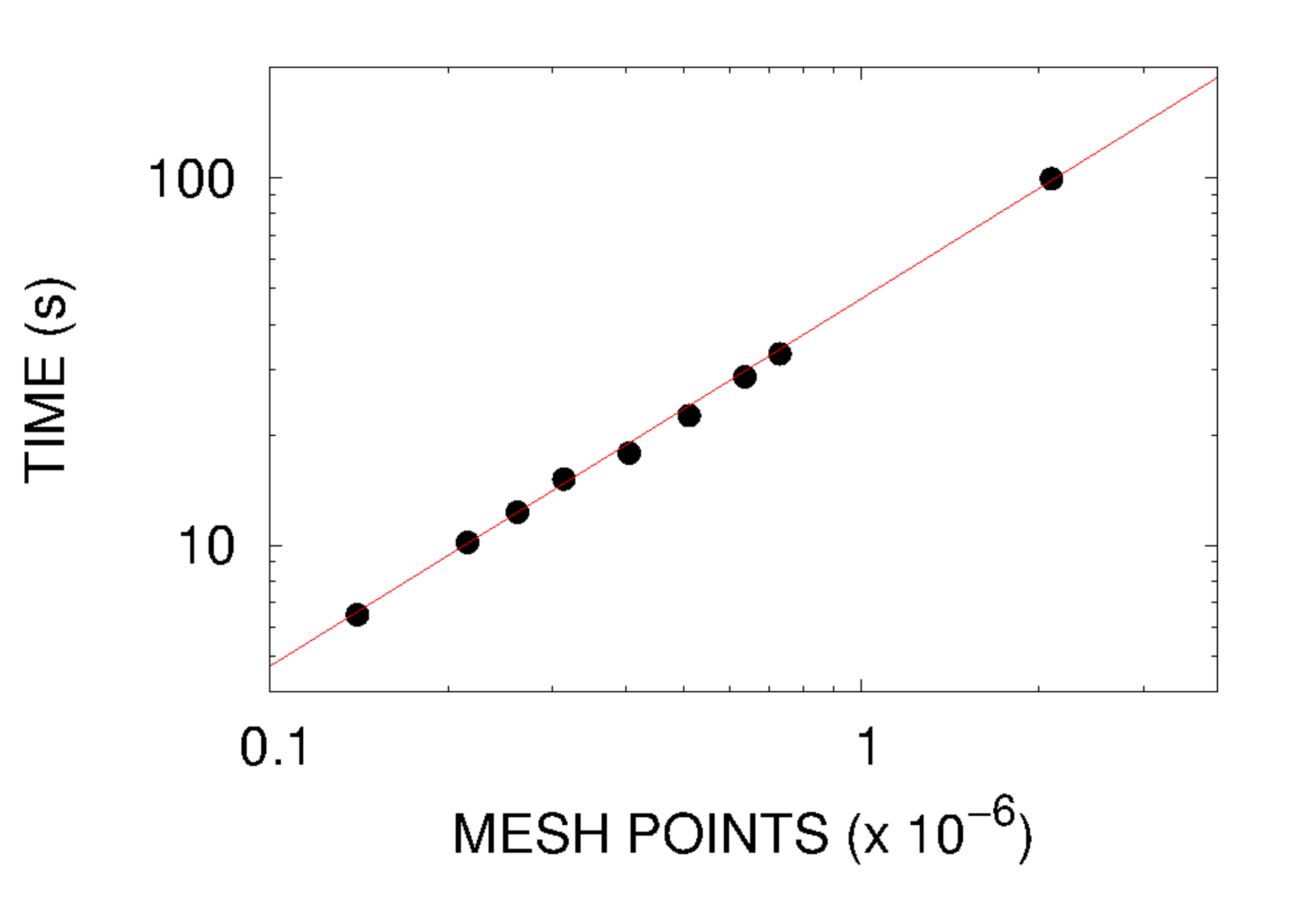} 
\end{center}
\caption{Log-log plot of the CPU time (sequential, Nehalem Intel Xeon X5550, 2.66 GHz) of the multi-grid solution of the Poisson equation  {\em vs}. system size. Line shows linear scaling. \label{complexityHypre}
}
\end{figure}

	Figure~\ref{complexityHypre} shows the linear dependence of the
sequential execution time of the MG solver on the size of the problem.

\section{Use of the O design atom}
\begin{figure}[hb]
\begin{center}
\includegraphics[scale=.27]{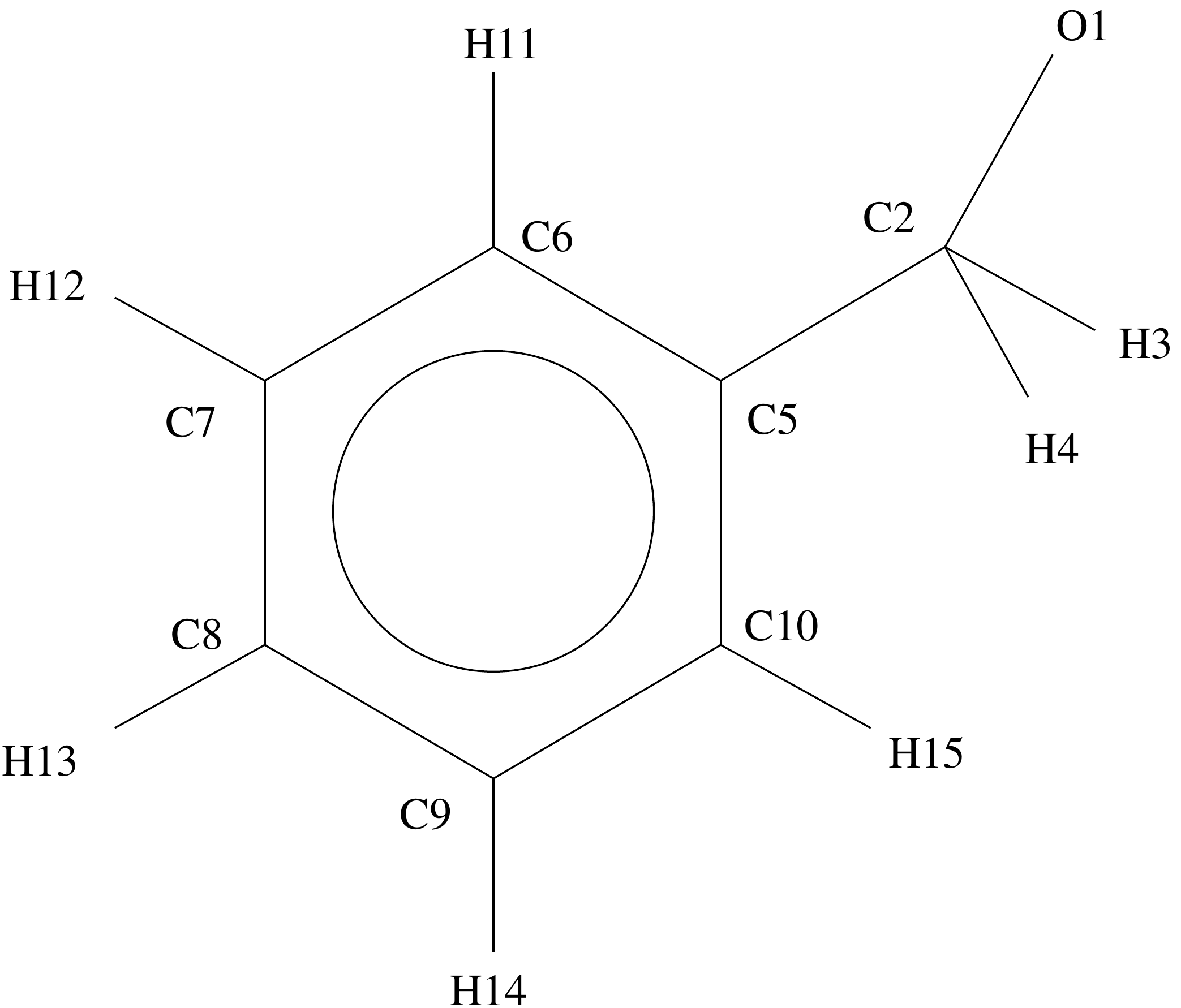} 
\end{center}
\caption{Atomic numbering used to compare the $Z$-matrices. \label{numbers}
}
\end{figure}
The $Z$-matrices of the optimised geometries (DZP, LDA, standard Troullier-Martins pseudo-potentials) of the full benzylcarbamide and the truncated form are provided below. Atom numbers differ from the main text; refer figure \ref{numbers} above.

\vspace{2cm}
\noindent{\bf $Z$-matrix full molecule}

\begin{verbatim}
 c
 c    1 cc2
 c    2 cc3         1 ccc3
 c    3 cc4         2 ccc4          1 dih4
 c    4 cc5         3 ccc5          2 dih5
 c    5 cc6         4 ccc6          3 dih6
 c    2 cc7         3 ccc7          4 dih7
 o    7 oc8         2 occ8          3 dih8
 h    7 hc9         2 hcc9          3 dih9
 h    7 hc10        2 hcc10         3 dih10
 h    3 hc11        2 hcc11         7 dih11
 h    4 hc12        3 hcc12         2 dih12
 h    5 hc13        4 hcc13         3 dih13
 h    6 hc14        5 hcc14         4 dih14
 h    1 hc15        2 hcc15         7 dih15

cc2         1.396646
cc3         1.391612
ccc3        119.687
cc4         1.395160
ccc4        119.692
dih4          0.093
cc5         1.392020
ccc5        120.426
dih5         -0.104
cc6         1.393252
ccc6        119.826
dih6          0.108
cc7         1.483357
ccc7        123.223
dih7        179.682
oc8         1.411371
occ8        112.767
dih8         -4.392
hc9         1.118685
hcc9        112.372
dih9        117.630
hc10        1.118928
hcc10       109.421
dih10      -124.994
hc11        1.106000
hcc11       117.974
dih11        -0.028
hc12        1.103650
hcc12       120.114
dih12      -179.891
hc13        1.103440
hcc13       120.430
dih13      -179.685
hc14        1.103566
hcc14       120.486
dih14      -179.925
hc15        1.105580
hcc15       119.052
dih15         0.187
\end{verbatim}

\vspace{2cm}
\noindent{\bf $Z-$-matrix Truncated molecule}

\begin{verbatim}
c
c    1 cc2
c    2 cc3         1 ccc3
c    3 cc4         2 ccc4          1 dih4
c    4 cc5         3 ccc5          2 dih5
c    5 cc6         4 ccc6          3 dih6
c    2 cc7         3 ccc7          4 dih7
o    7 oc8         2 occ8          3 dih8
h    7 hc9         2 hcc9          3 dih9
h    7 hc10        2 hcc10         3 dih10
h    3 hc11        2 hcc11         7 dih11
h    4 hc12        3 hcc12         2 dih12
h    5 hc13        4 hcc13         3 dih13
h    6 hc14        5 hcc14         4 dih14
h    1 hc15        2 hcc15         7 dih15

cc2         1.396569
cc3         1.392214
ccc3        119.781
cc4         1.392371
ccc4        120.107
dih4          0.046
cc5         1.391827
ccc5        120.063
dih5         -0.082
cc6         1.393444
ccc6        119.910
dih6          0.038
cc7         1.489138
ccc7        118.945
dih7       -179.458
oc8         1.429060
occ8        110.379
dih8         -4.063
hc9         1.117354
hcc9        110.988
dih9        114.752
hc10        1.117136
hcc10       111.543
dih10      -123.507
hc11        1.107117
hcc11       116.267
dih11         0.794
hc12        1.103764
hcc12       120.342
dih12      -179.991
hc13        1.103441
hcc13       120.126
dih13      -179.768
hc14        1.104104
hcc14       119.974
dih14      -179.783
hc15        1.105587
hcc15       120.026
dih15        -0.564 
\end{verbatim}

\begin{figure}[h]
\begin{center}
\includegraphics[scale=.27]{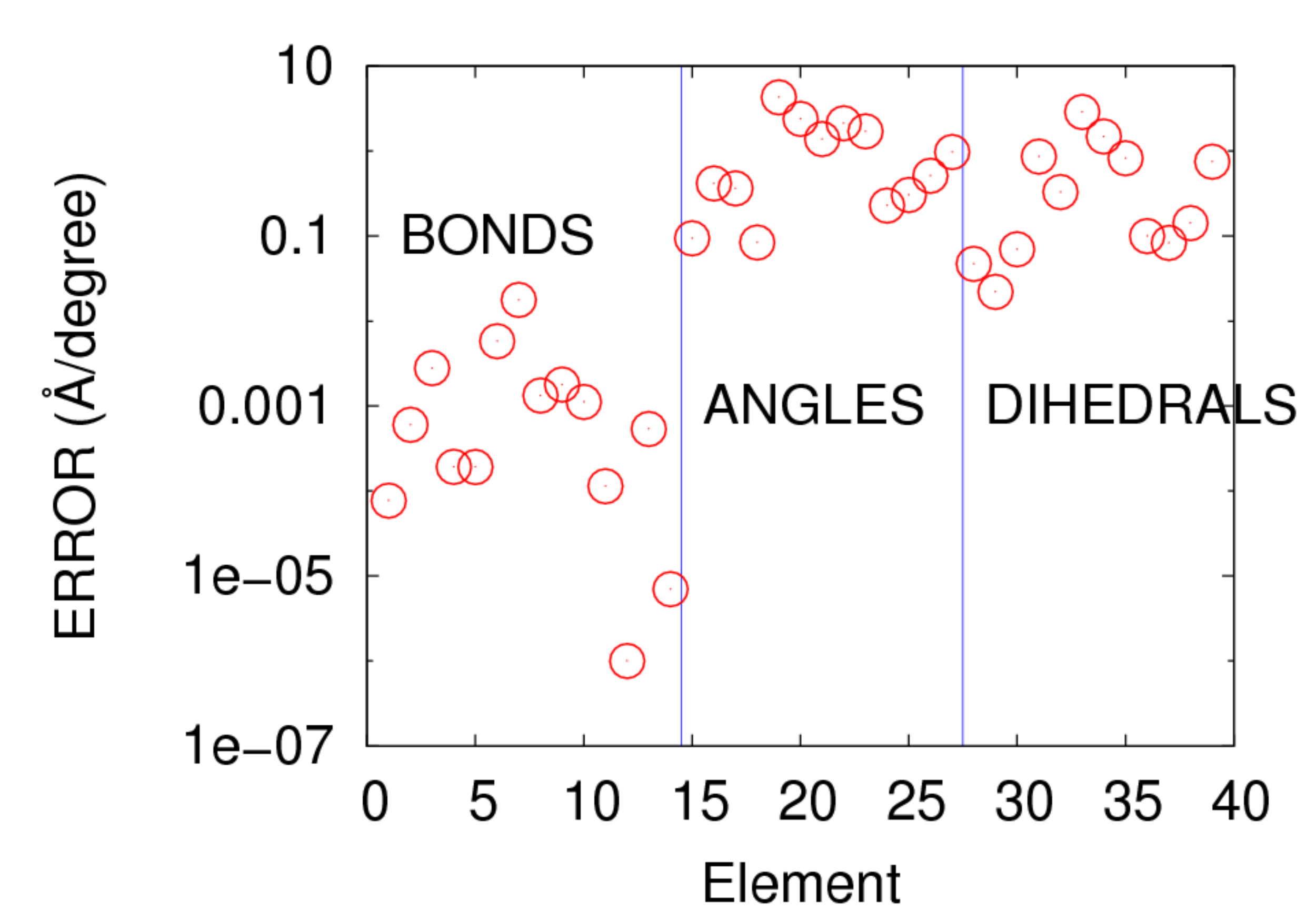} 
\end{center}
\caption{Errors of internal coordinates of the truncated {\em vs}. the full molecule (numbering, see $Z$-matrices and figure \ref{numbers}. \label{compare3}
}
\end{figure}

Figure \ref{compare3} provides an overview of the errors in the different types of internal coordinates, with reference to the $Z$-matrices and the atomic numbering in figure \ref{numbers} (not the same as in the main text).

\newpage
\section{Adaptive algorithm for fitting Lorentzians to the raw \tddft\ spectrum}
\noindent\begin{minipage}{\textwidth}
\centering
\begin{minipage}[c]{\linewidth}
  \centering
\begin{algorithm}[H]%
 \DontPrintSemicolon
 \KwSty{Initialization} ($p=1$): \;
 \Begin{Start with  a small number $N_1$ of uniformly distributed frequency points\;
$$\omega_j^1=\omega_{\tsub{min}} + \left( \omega_{\tsub{max}} -
\omega_{\tsub{min}}\right) \times \frac{j-1}{N_p}$$ and a not too small value
of $\epsilon = \epsilon_1 $. \;
}
\For{it = 1, iterMax}{
Compute \textit{P} on the current $\omega_j$.\;
\KwSty{Build the new distribution:}\;
\Begin{
 Define a set of uniformly spaced $U_i = i/N_p , i=1,\ldots N_p $ in [0,1].\;
	Find $\omega_i$ such that $J(\omega_i) \leq U_i <
          J(\omega_{i+1})$ \;
          The new improved frequency is $\omega_i =
          J_{\hbox{\scriptsize approx}}^{-1}(U_i)$
		}
		\nlset{step Fit}\label{InAlgoFit}Perform the least squares fit and determine the transitions $\Omega_k$ and  associated oscillator strengths $f_k$.\;
		Check the convergence: \;
		\hspace{10mm}$\displaystyle\max{(\max_{k < N_L}{\|\Omega_k^{it}- \Omega_k^{it-1}\|},
		\max_{k < N_L}{\|f_k^{it}- f_k^{it-1}\|})} < \mu$ \;
		
		Reduce the regularization parameter if necessary.  \;
}
\caption{Adaptive algorithm}
\label{algo}
\end{algorithm}
\end{minipage}
\end{minipage}

The different steps of the adaptive algorithm are illustrated in
Algorithm~\ref{algo}.  We start the algorithm with a uniform distribution of points.
In \ref{InAlgoFit} of Algorithm~\ref{algo}, first we determine the number of Lorentzian involved in the spectra, using a peak detection algorithm, inspecting for  adjacent regions of increasing and decreasing values. The adaptive algorithm converges when the maximum difference between two iterations of the $N_{\tsub{trans}}$ first (\ie\ strongest) transition frequencies and the oscillator strengths are less than a given threshold $\mu$.

Tables \ref{Tab:eps} and \ref{Tab:minN} below provide additional data to complement the main text, on the amount of information recovered under different conditions in the spectrum of indigo in the range $[0.02, 0.4]$\,Ry\,. They show respectively, the influence of the regularisation parameter $\epsilon$ at constant number of data points $N_{\tsub{freq}}=257$ and at minimal $N_{\tsub{freq}}= 1 + \frac{3}{2\epsilon}( \omega_{max}- \omega_{min})$. In both cases, the data points are distributed \textit{uniformly}. 
The residual $\chi^2$  is defined by 
\begin{equation*}
\chi^2 =  |( I(\omega)- \chi''_{\tsub{Lorentzian}}(\omega) )|_2/|I(\omega)|_2 ,
\end{equation*}
\noindent where $|.|_2$ is the discrete $l^2$ norm, $\chi''_{\tsub{Lorentian}}$ is the Lorentian approximation to the experimental data $I(\omega)$. All other calculation conditions are the same as in the main text.

\begin{table}[htbp]
\centering\begin{tabular}{|l|c|r|c|c|c|c|c|c|c|}
\hline
                  &Transitions        &  \multicolumn{2}{|c|}{First transition} & \multicolumn{2}{|c|}{Second transition} &  \\ 
 $\epsilon$ (Ry)  &recovered,         & $\Omega_1$ (Ry)   &       $f_1$         & $\Omega_2 (Ry)$    &  $f_2$             & $\chi^2$\\ 
                  &$N_{\tsub{trans}}$ &                   &                     &                    &                    &\\ \hline    
0.5     &      0 &   -       &   -           &   -           &   -   & 0.45\\
0.1     &      1 & 0.322242  &     0.129389  &   -           &   -   & 0.48\\
0.075   &      1 & 0.318569  &     0.104083  &   -           &   -   & 0.48\\
0.05    &      2 & 0.318730  &     0.103203  & 0.163087   & 0.180037 & 0.62\\
0.025   &      2 & 0.313096  &     0.857635  & 0.153033   & 0.160789 & 0.15\\
0.01    &      5 & 0.312138  &     0.814819  & 0.151514   & 0.194918 & $5.9\times 10^{-2}$\\
0.0075  &      6 & 0.312170  &     0.812227  & 0.151501   & 0.196285 & $5.5\times 10^{-2}$\\
0.005   &      7 & 0.312170  &     0.810278  & 0.151497   & 0.196530 & $4.2\times 10^{-2}$\\
0.0025  &      7 & 0.312168  &     0.809110  & 0.151496   & 0.196621 & $4.1\times 10^{-2}$\\
0.001   &     10 & 0.312168  &     0.809302  & 0.151496   & 0.196691 & $2.4\times 10^{-3}$\\
0.00075 &     12 & 0.312168  &     0.809151  & 0.151496   & 0.196687 & $1.9,10^{-3}$\\
0.0005  &     10 & 0.312168  &     0.809377  & 0.151496   & 0.196662 & $3.4\times 10^{-3}$\\
0.00025 &      7 & 0.312169  &     0.807611  & 0.151497   & 0.196209 & $3.8\times 10^{-3}$\\
0.0001  &     13 & 0.310937  &     *            &     *         & *&$\infty$\ \\\hline
\end{tabular}
\caption{Influence of $\epsilon$ (at  $N_{\tsub{freq}} = 257$) on the frequency and oscillator strength recovered for the two strongest transitions of indigo in the interval $[0.02, 0.4]$\,Ry\,. - means nothing found, * means wrong resultss (negative polarisability).}\label{Tab:eps}
\end{table}

\begin{table}[hb]
\centering\begin{tabular}{|l|c|c|c|c|c|c|c|}
\hline
                &                    & Transitions               &  \multicolumn{2}{|c|}{First transition} & \multicolumn{2}{|c|}{Second transition} &   \\
$\epsilon$ (Ry) &  $N_{\tsub{freq}}$ & recovered,                & $\Omega_1 (Ry)$      &       $f_1$      & $\Omega_2 (Ry)$   &  $f_2$              &  $\chi^2$ \\  
                &                     &  $N_{\tsub{trans}}$      &                      &                  &                   &                     & \\ \hline  
0.1     &   6&  0 &     -        &         -        &   -           &   -  & 0.48\\
0.075   &   8&  0 &     -        &         -        &   -           &   -  &0.48\\
0.05    &  12&  0 &     -        &        -         &   -           &   - & 0.62\\
0.025   &  23&  2 & 0.312846  &     0.876805  & 0.153001   &  0.173854 & 0.14\\
0.01    &  58&  4 & 0.312145  &     0.813828  & 0.151517   &  0.193658 & $5.9 \, 10^{-2}$\\
0.0075  &  77&  4 & 0.312159  &     0.809740  & 0.151502   &  0.193412 & $5.5 \, 10^{-2}$\\
0.005   & 115&  6 & 0.312170  &     0.808326  & 0.151497   &  0.195282 & $4.2 \, 10^{-2}$\\
0.0025  & 229&  7 & 0.312168  &     0.809217  & 0.151496   &  0.196632 & $4.0 \, 10^{-2}$\\
0.001   & 571& 11 & 0.310168  &     0.804095  & 0.149496   &  0.194092 & $2.5 \, 10^{-3}$\\
0.00075 & 761& 13 & 0.309168  &     0.801070  & 0.148496   &  0.192807 & $2.3 \, 10^{-3}$\\
0.0005  &1141& 16 & 0.308168  &     0.798482  & 0.148830   &  0.193213 & $1.7 \, 10^{-3}$\\
0.00025 &2281& 23 & 0.307502  &     0.794531  & 0.149496   &  0.193877 & $3.4 \, 10^{-3}$ \\
0.0001  &5701& 22 & 0.304964  &     0.726735  & 0.148896   &  0.191774 & $\infty$\\\hline
\end{tabular}
\caption{Influence of $\epsilon$, automatic choice for the number of frequencies ( - means nothing found.)}
\label{Tab:minN}
\end{table}

\end{document}